\documentclass[pre,aps,twocolumn,showpacs,floatfix]{revtex4-1}

\usepackage{dcolumn}
\usepackage{amsmath}
\usepackage{amssymb}
\usepackage{bm}
\usepackage{graphicx}
\usepackage{epsf,epsfig}
\usepackage{hyperref}
\usepackage{color}

\begin{document}
\title{Frictional rigidity percolation and minimal rigidity proliferation: From a new universality class to superuniversality}
\author{Kuang Liu$^1$, S. Henkes$^2$, and J. M. Schwarz$^1$}
\affiliation{$^1$ Department of Physics and Syracuse Soft and Living Matter Program, Syracuse University, Syracuse, NY USA\\
$^2$ Institute of Complex Systems and Mathematical Biology, Department of Physics, University of Aberdeen, Aberdeen, Scotland, United Kingdom}
\begin{abstract}
We introduce two new concepts, frictional rigidity percolation and minimal rigidity proliferation, to help identify the nature of the frictional jamming transition as well as significantly broaden the scope of rigidity percolation. For frictional rigidity percolation, we construct rigid clusters in two different lattice models using a $(3,3)$ pebble game, while taking into account contacts below and at the Coulomb threshold. The first lattice is a honeycomb lattice with next-nearest neighbors, the second, a hierarchical lattice. For both, we generally find a continuous rigidity transition.  Our numerical results suggest that, for the honeycomb lattice, the exponents associated with the transition found with the frictional $(3,3)$ pebble game are distinct from those of a central-force $(2,3)$ pebble game.  We propose that localized motifs, such as hinges, connecting rigid clusters that are allowed only with friction could give rise to this new frictional universality class. However, the closeness of the order parameter exponent between the two cases hints at potential superuniversality. To explore this possibility, we construct a bespoke cluster generating algorithm invoking generalized Henneberg moves, dubbed minimal rigidity proliferation. The minimally rigid clusters the algorithm generates appear to be in the same universality class as connectivity percolation, suggesting superuniversality between all three types of transitions. Finally, the hierarchical lattice is analytically tractable and we find that the exponents depend both on the type of force and on the fraction of contacts at the Coulomb threshold. These combined results allow us to compare two universality classes on the same lattice via rigid clusters for the first time to highlight unifying and distinguishing concepts within the set of all possible rigidity transitions in disordered systems.

\end{abstract}
\maketitle
\section{Introduction}
At the heart of every rigidity transition is the emergence of a spanning rigid cluster---a cluster in which the bonds of an underlying contact network, be it particles or springs, are rigid with respect to each other. For disordered systems, the starting point of choice has become randomly-diluted spring networks with central-force interactions\cite{Feng1984,Feng1985,Tremblay1986,Day1986}. For example, as bonds are randomly diluted from a triangular lattice, be it regular or one in which the lattice points have been slightly randomized, i.e. generic, the system goes from rigid with a non-zero shear modulus to floppy with zero shear modulus~\cite{Roux1988,Jacobs1995,Jacobs1996,Moukarzel1995}. Underlying this mechanical phase change is the transition from a system with a spanning rigid cluster to a system with no spanning rigid cluster, as identified by the combinatorial (2,3) pebble game~\cite{Jacobs1997}. The location of this rigidity transition approximately occurs where the number of degrees of freedom are frozen out by the number of constraints--the number of force-balance equations, otherwise known as Maxwell constraint counting~\cite{Maxwell1864}. 

The nature of the rigidity transition in the central-force, randomly bond-diluted triangular lattice is a continuous one with a correlation length exponent, $\nu=1.21\pm 0.06$, an order parameter exponent $\beta=0.18\pm 0.02$, and a fractal dimension of the spanning rigid cluster, $d_f=1.86 \pm 0.02$~\cite{Jacobs1995,Jacobs1996}.  Note that while the lattice in these studies was a generic one, these exponents were obtained via the combinational pebble game and so are independent of whether or not the underlying triangular lattice is regular or generic, unlike elastic exponents or other properties extracted from, say, the dynamical matrix~\cite{Roux1988,Jacobs1995}. From now on, we will assume any underlying lattice is regular unless otherwise specified. The exponents are slightly different from connectivity percolation transition with two-dimensional exponents $\nu=4/3$, $\beta=5/36$, and $d_f=91/48$~\cite{Nienhuis1987} Despite the small difference in exponents, it was eventually argued that the rigidity percolation transition is in a different universality class from connectivity percolation since there are nonlocal interactions and rigidity is a vector problem given that forces are vectors, unlike connectivity percolation~\cite{Jacobs1995,Jacobs1996}. 

On the other hand, lattices with no loops, i.e. Bethe lattices, are amenable to analytical treatment and demonstrate that the spanning rigid cluster at the transition is not fractal~\cite{Moukarzel1997,Moukarzel1999}.  To add to the complexity, numerical simulations of three-dimensional lattices with central-force interactions indicate a discontinuous rigidity transition as well, in contrast to the two-dimensional case~\cite{Chubynsky2007}.  Finally, central-force models with next-neighbor springs can exhibit hybrid rigidity transitions exhibiting both continuous and discontinuous features~\cite{Zhang2015}. With this rather varied set of phase transition outcomes depending just on the type of lattice, a general solution to the rigidity percolation problem is not yet clear, if it is at all possible. 

Rigidity percolation with bond-bending forces has also been studied, thereby adding yet another dimension to the problem~\cite{Kantor1984,Feng1985b,Schwartz1985,Schwartz1985,Wang1996,Das2012}.  Numerical simulations of two-dimensional systems measuring elastic properties suggest that bond-bending forces drive the transition into a different universality class~\cite{Head2003}. However, since the pebble game has not yet been applied to bond-bending forces even in two-dimensions, a direct comparison between bond-bending and central-force rigidity percolation in terms of the order parameter exponent and the correlation length exponent has yet to be even made. In other words, how the nature of the rigidity transition depends on the type of force is even less understood than the type of lattice. 

Particle packings also undergo a rigidity transition as a function of the packing fraction~\cite{Durian1995,Ohern2003,Majumdar2007} and possess the additional feature of contact network rearrangements that is not allowed in a randomly-diluted spring network.  The rigidity transition in such systems has been typically labeled a jamming transition. Nonetheless, the transition is from a zero to non-zero bulk modulus as the packing fraction increases, suggesting the emergence of a spanning rigid cluster~\cite{Ohern2003}. This suggestion was made explicit by extracting the contact network of a two-dimensional {\it frictionless} particle packing at the jamming transition, where there are only central forces, and constructing the rigid clusters from this network via the (2,3) pebble game to find that at the onset of rigidity/jamming every particle participating in the contact network is part of one rigid cluster, i.e. the spanning rigid cluster is bulky at the transition~\cite{Ellenbroek2015}. 

\begin{figure*}[t]
\includegraphics[width=0.99\textwidth]{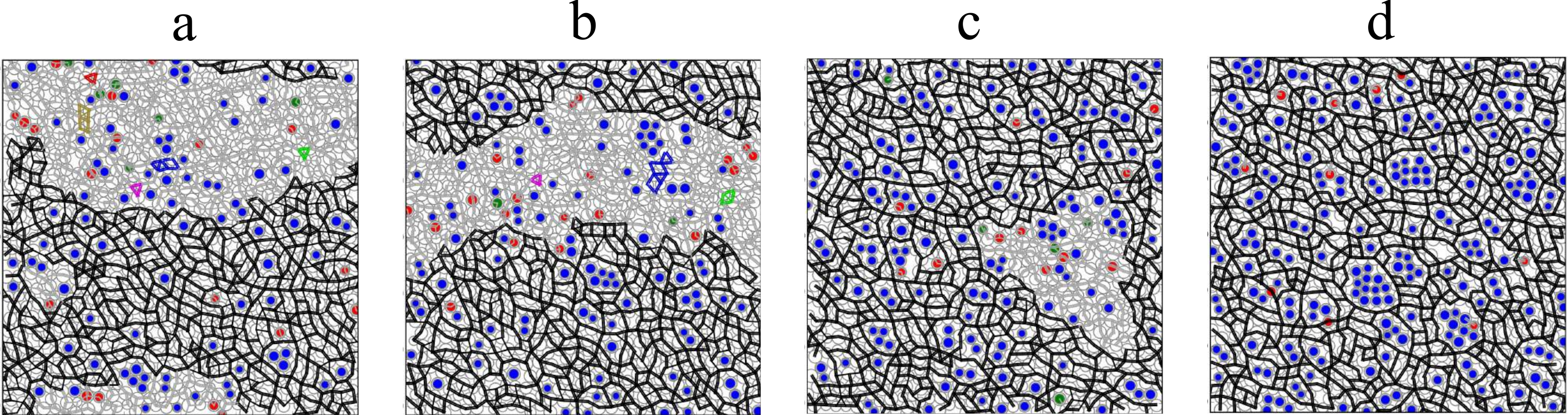}\label{fig:rigidclusters}
\caption{Four snapshots from the molecular dynamics simulation showing partically rigid systems close to the frictional rigidity transition. In black is the largest rigid cluster, while floppy regions are colored gray. Blue, green and red disks correspond to three, two and one leftover pebble, respectively.
The range of partial rigidity decreases with increasing friction coefficient, or equivalently, increasing $q$. (a) $\mu=0.2$, with a transition at $q=0.78$ and $z=3.35$. (b) $\mu=0.3$, with a transition at $q=0.86$ and $z=3.15$, (c) $\mu=0.5$, with a transition at $q=0.95$ and $z=3.0$, and finally (d) $\mu=10$ with a transition at $q=1.0$ and $z=2.8$. The last value is due to the large number of contactless particles (rattlers) in the packing, visible in blue. }
\label{fig:rigidclusters1}
\end{figure*}

If one were to turn to {\it frictional} systems, how does one compute rigid clusters from networks abstracted from two-dimensional frictional particle packings?  A 2016 PRL by two of us outlined a new pebble game algorithm for doing so for both translational and rotational degrees of freedom~\cite{Henkes2016} and applied it to molecular dynamics simulations of frictional particle packings at a fixed packing fraction experiencing slow shear.  As the packing goes from jammed to unjammed and back again, rigid clusters were determined at constant strain intervals. The size of the largest rigid cluster indicated a continuous transition, and so did the observation of a roughly power-law distribution of rigid cluster sizes near the jamming transition~\cite{Henkes2016}.  Within the spanning rigid clusters, regions of floppiness were found, suggesting partial rigidity. These floppy regions are reminiscent of the rigid clusters found in the canonical bond-diluted triangular lattice. The floppy regions also appeared to be physically relevant as the non-affine motion of the particles was larger outside the rigid cluster as compared to inside.  We present four such rigid cluster images close to the transition for four different values of the friction coefficient in Fig.~\ref{fig:rigidclusters1}.

One of the most natural questions arising from this recent study of frictional rigid clusters is whether or not the rigidity transition is actually a continuous one with regards to the onset of the spanning rigid cluster. A continuous rigidity transition would be very different from the frictionless case where there are only repulsive, central forces. Since the molecular dynamics simulation performed earlier does not allow one to tune the system to be arbitrarily close to the rigidity transition, it is difficult to assess the nature of the onset of the spanning rigid cluster using such simulations.

In this article, we construct a lattice model of a frictional packing based on a honeycomb lattice with next-nearest neighbors. We denote some bonds as below the Coulomb threshold (frictional) and some bonds as at the Coulomb threshold (sliding) and implement the frictional (3,3) pebble game to construct rigid clusters and study the nature of the rigidity transition, now in the context of frictional rigidity percolation to broaden the applicability of rigidity percolation. 
Finite-size-scaling provides strong evidence that the frictional percolation transition is indeed a continuous transition, with the exception of some limiting cases. The exponents that we compute numerically are distinct from those found for central-force rigidity percolation, therefore signifying a change in universality class between the two cases. We focus on rigid hinges and other one-dimensional connectors as mechanisms for propagating rigidity that are different between the two cases. However, both the order parameter exponents and the fractal dimension are statistically similar across all our countinuous transitions. This is a hint of superuniversality, which is usually considered in the context of symmetry classes in ordinary continuous phase transitions driven by symmetry breaking, even though here, there is no apparent symmetry that is explicitly broken. Our numerical results are summarized in Fig.~\ref{fig:exponents}. 

We support our finite-size scaling analysis near the transition by algorithmically exploring the possibility of superuniversality within a subset of configurations, namely those that are minimally rigid.  By (1) using concepts from invasion percolation~\cite{Wilkinson1983}, used to build only spanning clusters in connectivity percolation, and (2) extending the Henneberg moves~\cite{Henneberg1911}, which are used to grow a minimally rigid network with central-forces only, we contruct a new algorithm to grow a minimally rigid network with frictional forces, which we dub minimal rigidity proliferation. Through this, we demonstrate that there is indeed a way to construct minimally rigid spanning clusters whose structure is the same between frictional rigidity percolation, central-force rigidity percolation, and even connectivity percolation, making the case for superuniversality, at least for such clusters. 

While the honeycomb lattice approach is numerical, we also employ analytical techniques to study the frictional rigidity transition in hierarchical lattices, and compare these results with our honeycomb lattice. The hierarchical lattice solution expands the number of analytical results in the field of rigidity percolation. We conclude with a discussion about the onset of rigidity in frictional particle packings as compared to frictionless ones, investigate how our two-dimensional results connect (or do not connect) with three-dimensional results, and finally, how our results impact the field of rigidity percolation more broadly. 

\section{Honeycomb lattice with next-nearest neighbors}

\subsection{Model}

\begin{figure}[t]
\includegraphics[width=0.99\columnwidth]{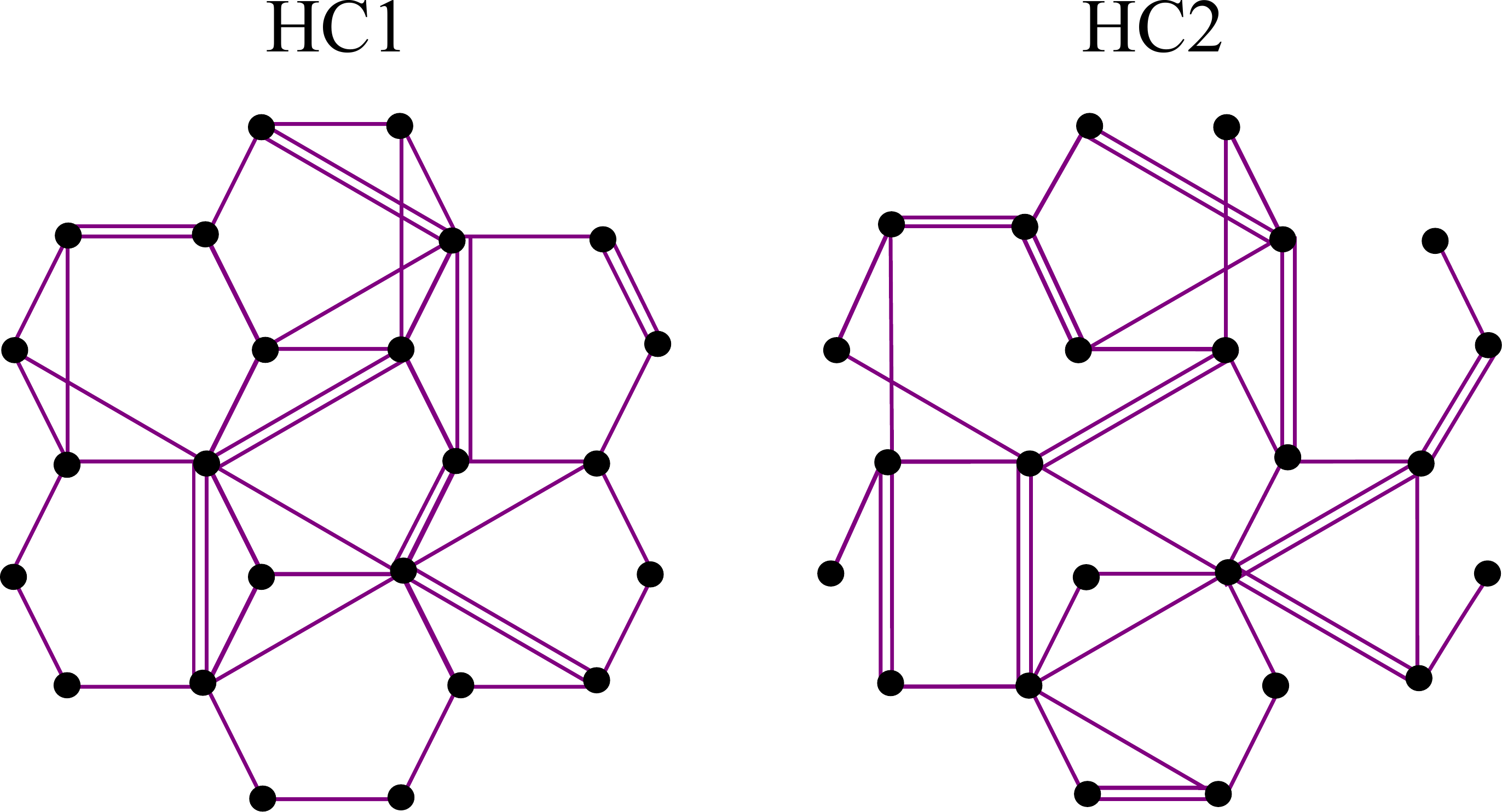}\label{fig:examplemodel}
\caption{Schematic models HC1 and HC2. These random networks are constructed by either adding next-nearest neighbor (NNN) bonds to an occupied honeycomb lattice (HC1) or adding random first and second neighbor bonds (HC2), all with probability $p$. Double bonds denote frictional/gear-like bonds, which occur with probability $q$, while single bonds denote sliding bonds.}
\end{figure}

To motivate the model we begin with by reviewing Maxwell constraint counting in two-dimensional frictionless packings with $N$ particles and average coordination number $z$~\cite{Maxwell1864}. The total number of degrees of freedom is $2N$, while there are  $\frac{N_sz}{2}$ force-balance constraints. When the number of degrees of freedom is equal to the number of force-balance constraints, the system is minimally rigid, i.e. 
\begin{equation} 
\label{eq:cons_count} 
2N-3=\frac{z_c N}{2}, 
\end{equation}
where the number of global rigid body translations and rotations has been subtracted out since they are trivial. This equation yields the critical average coordination number \(z_c=4\) (as $N\rightarrow \infty$) for the onset of rigidity. Much numerical work with frictionless particle packings has shown that this counting is an extremely good method to determine the rigidity transition point~\cite{VanHecke2009,Liu2010,Wyart2010}.  No states of self-stress are observed in particle packings at the transition, since the values of the purely repulsive forces are uniquely determined by the boundary conditions at this point, such that a more involved constraint counting approach is not needed~\cite{Laman1979}. 

\begin{figure}[h]
    \includegraphics[width=0.85\columnwidth]{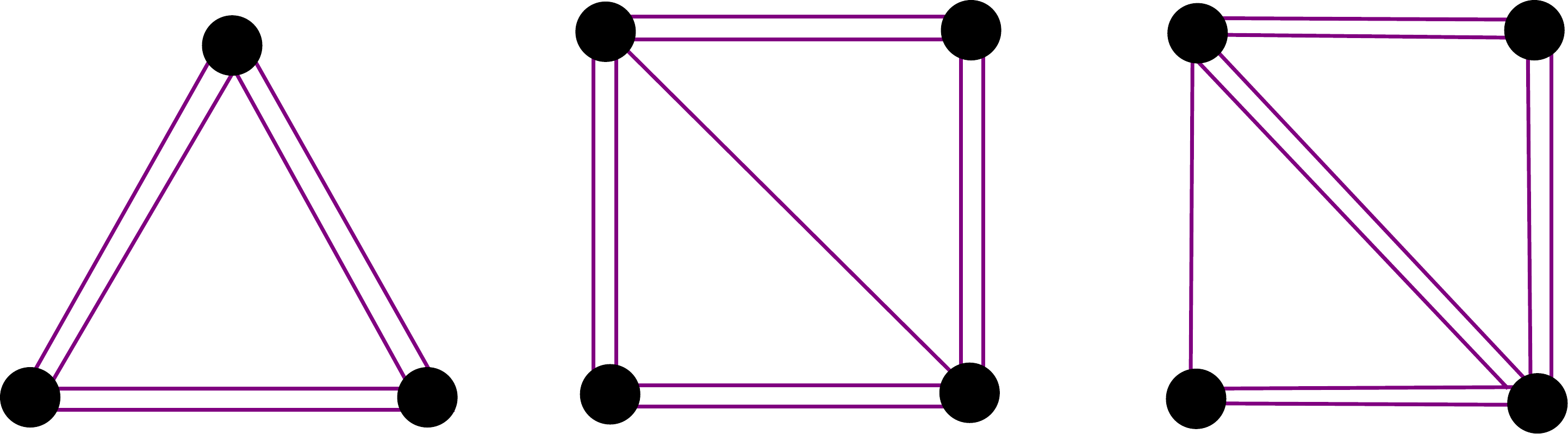}
    \caption{For a triangular constraint network with all double bonds, then this network is minimally rigid via the $(3,3)$ pebble game. For a four-site constraint network, there are 5 possible configurations in which this network is minimally rigid via the $(3,3)$  pebble game, two of which are presented.}
    \label{fig:example33pebble}
\end{figure}

For two-dimensional frictional particle packings, some contacts are below the Coulomb threshold with the magnitude of the tangential force less than the magnitude of the repulsive, central force times the friction coefficient. At such contacts, two particles can only rotate and translate with respect to one another just as a gear does, and these are denoted as frictional contacts. There are also contacts at the Coulomb threshold in which two particles slide with respect to each other. For these sliding contacts, the magnitude of the tangential force is set by the magnitude of the repulsive, central force, i.e. there is only one constraint. We distinquish between these two types of contacts by denoting $q$ to be the probability of having a frictional contact with $1-q$ denoting the probability of having a sliding constact, i.e. if $q=1$, all contacts are frictional. Then, performing the Maxwell constraint counting as above, since each particle has 3 translational and rotational degrees of freedom, there are $3(N-1)$ total degrees of freedom (subtracting out the trivial global degrees of freedom in which there is no relative motion between the particles).  Moreover, the interparticle forces yield $\frac{z(1+q)N}{2}$ total constraints.  We, therefore, arrive at the minimal rigidity criterion, or  
\begin{equation} 
\label{eq:gen_ziso}
3N-3=\frac{(1+q)z_cN}{2},
\end{equation}
where $q$ denotes the probability of having a frictional bond. For $N\rightarrow \infty$ and $q=1$, then all bonds are frictional and $z_c=3$.  If $q=1/2$, then $z_c=4$. Therefore, Eq. \ref{eq:gen_ziso} describes a \emph{line} of transition points interpolating from $z_c=3$ to $z_c=4$ as the ratio of frictional to sliding bonds changes. This method of counting is now known as generalized isostaticity~\cite{n_m,n_m2}. Such bounds are indeed observed in experiments~\cite{Onoda}. Note that increasing $\mu$ increases $q$, and in addition, that one cannot smoothly interpolate between frictional and frictionless packings as one cannot smoothly interpolate between 2 and 3 degrees of freedom. 

\begin{figure*}[t]
\includegraphics[width=0.7\textwidth]{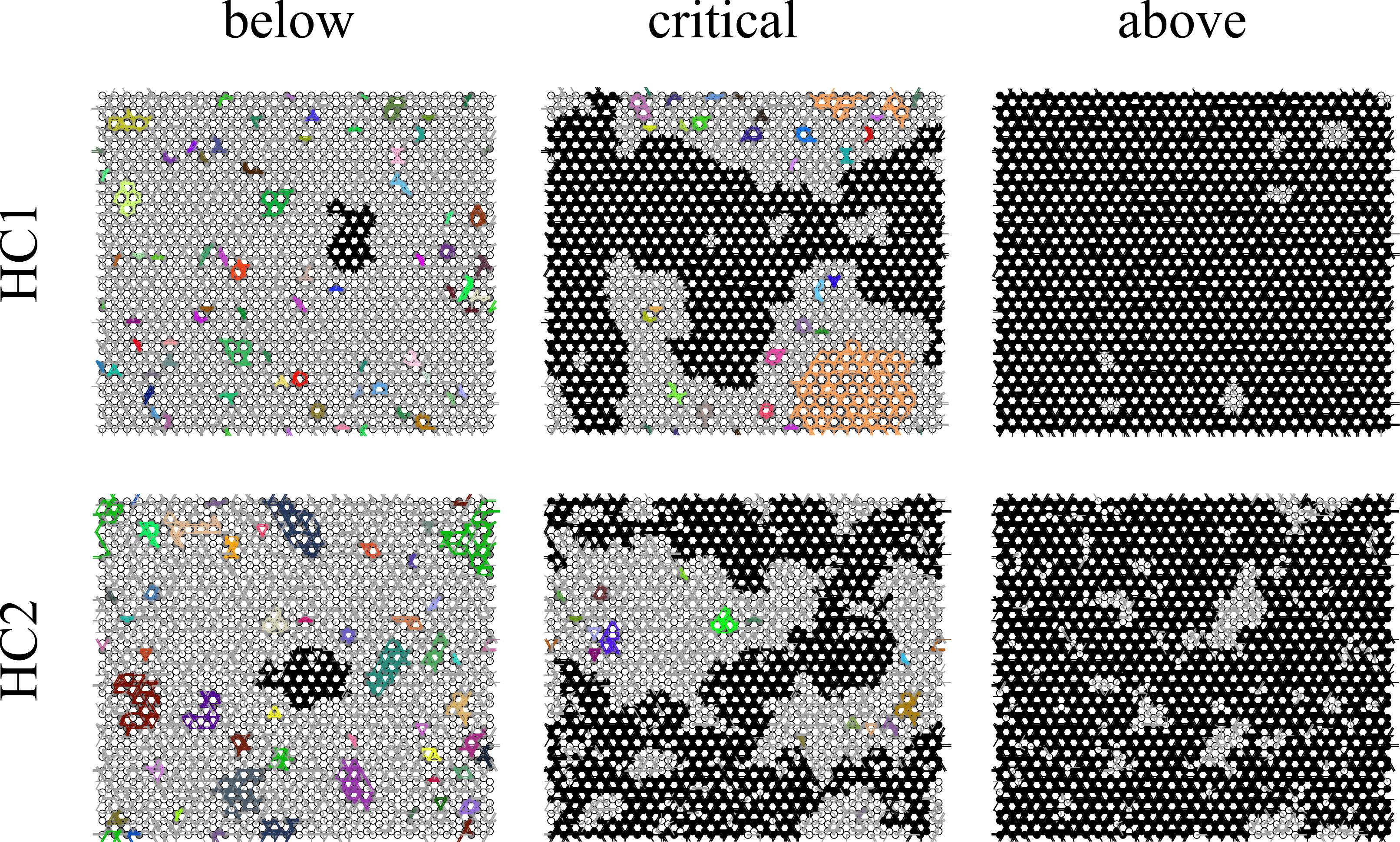}
\caption{Rigid clusters below, at, and above the rigidity transition for $HC1$ and $HC2$ in the (3,3) pebble game with $q=0.5$. Rigid clusters are colored, with the largest cluster in black, while floppy regions are in grey.}
\label{fig:rigidclusters2}
\end{figure*}

\begin{figure*}
\includegraphics[width=0.7\textwidth]{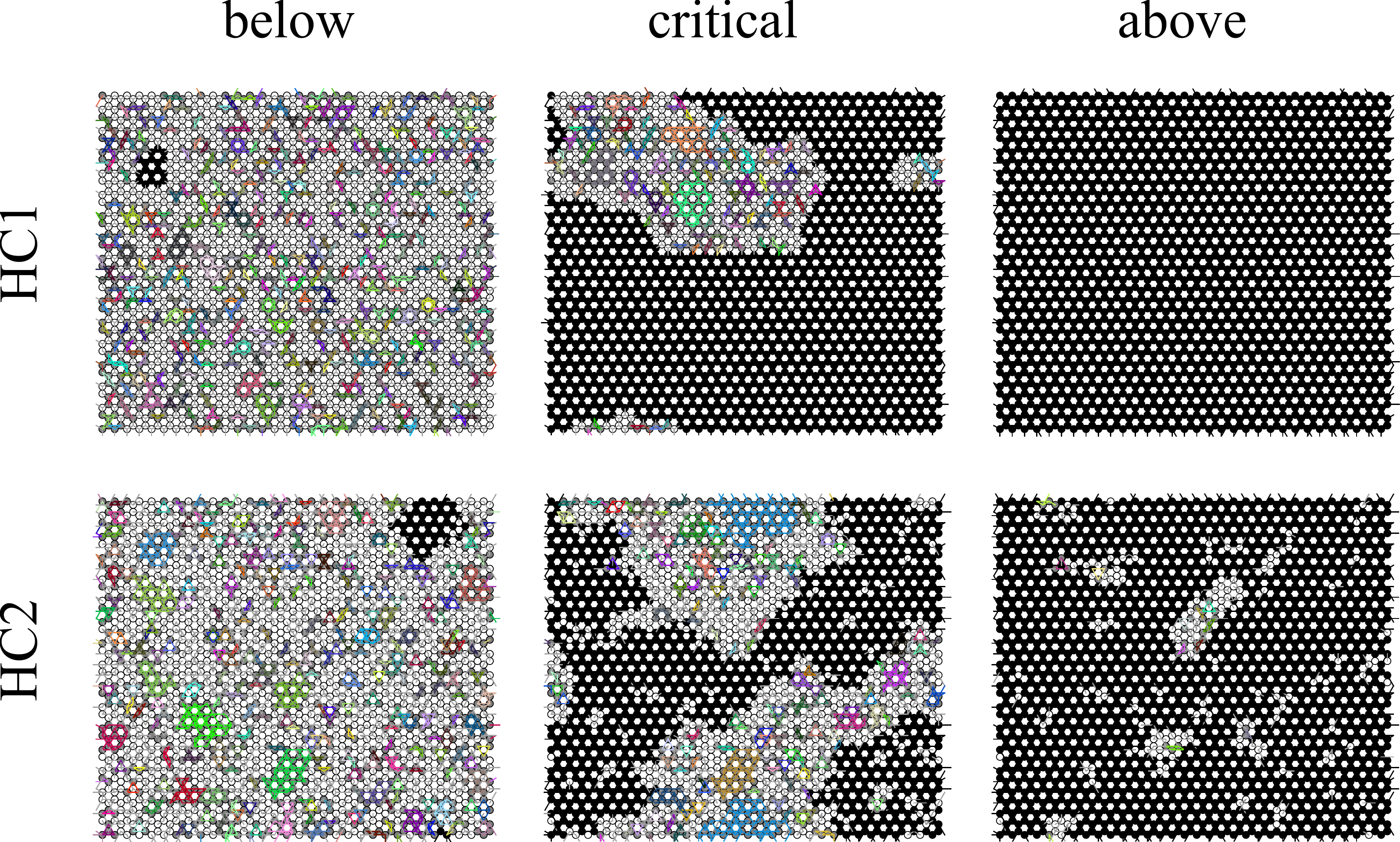}
\caption{Rigid clusters below, at, and above the rigidity transition for $HC1$ and $HC2$ for the $(2,3)$ pebble game. The black indicates the largest rigid cluster, and floppy regions are gray again. }
\label{fig:rigidclusters3}
\end{figure*}

To construct a lattice model for frictional particle packings, we consider a honeycomb lattice with additional next-nearest neighbor (NNN) bonds.  This modified honeycomb lattice has a maximum coordination number of $z_{max}=9$.  We define $p$ as the probability of bond/contact occupation.  The reason we employ the honeycomb lattice with next-nearest neighbors is because we can explore geometry to determine whether or not it is relevant for determining the nature of the phase transition.  We do so by constructing and studying two different models for bond occupation, see Figure~\ref{fig:examplemodel}. For the first model, we fully occupy the honeycomb backbone such that $p=1/3$ and then occupy the additional NNN bonds occupied randomly such that $p\ge 1/3$. We dub this first strategy of bond occupation $HC1$. We also implement a second strategy of bond occupation in which the bonds, both nearest-neighbor or next-nearest-neighor, are occupied at random, which we dub $HC2$.  For $HC1$, since the honeycomb backbone is fully occupied, the central forces on each particle can be balanced, which is required by local mechanical stability in frictionless, but not frictional, packings. Therefore, $HC1$ will allow us to more readily compare with the geometry of frictionless packings in order to see how frictionless differs from frictional.  The frictional bonds are then randomly assigned with probability $q$. Periodic boundary conditions are implemented.

Now we address the frictional versus sliding bonds for this lattice model. Since frictional bonds randomly occur with probability $q$, using Eq. \ref{eq:gen_ziso}  the critical occupation probability predicted by Maxell constraint counting is  
\begin{equation}
p_c=\frac{z_c}{z_{max}}=\frac{2}{3(1+q)}.
\label{eq:pcequation}
\end{equation}
Equation~\ref{eq:pcequation} tells us how $p_c$ depends on $q$, therefore, denoting a phase transition line between floppy and rigid phases just as in generalized isostaticity.  

\begin{figure*}[t]
\includegraphics[width=0.99\textwidth]{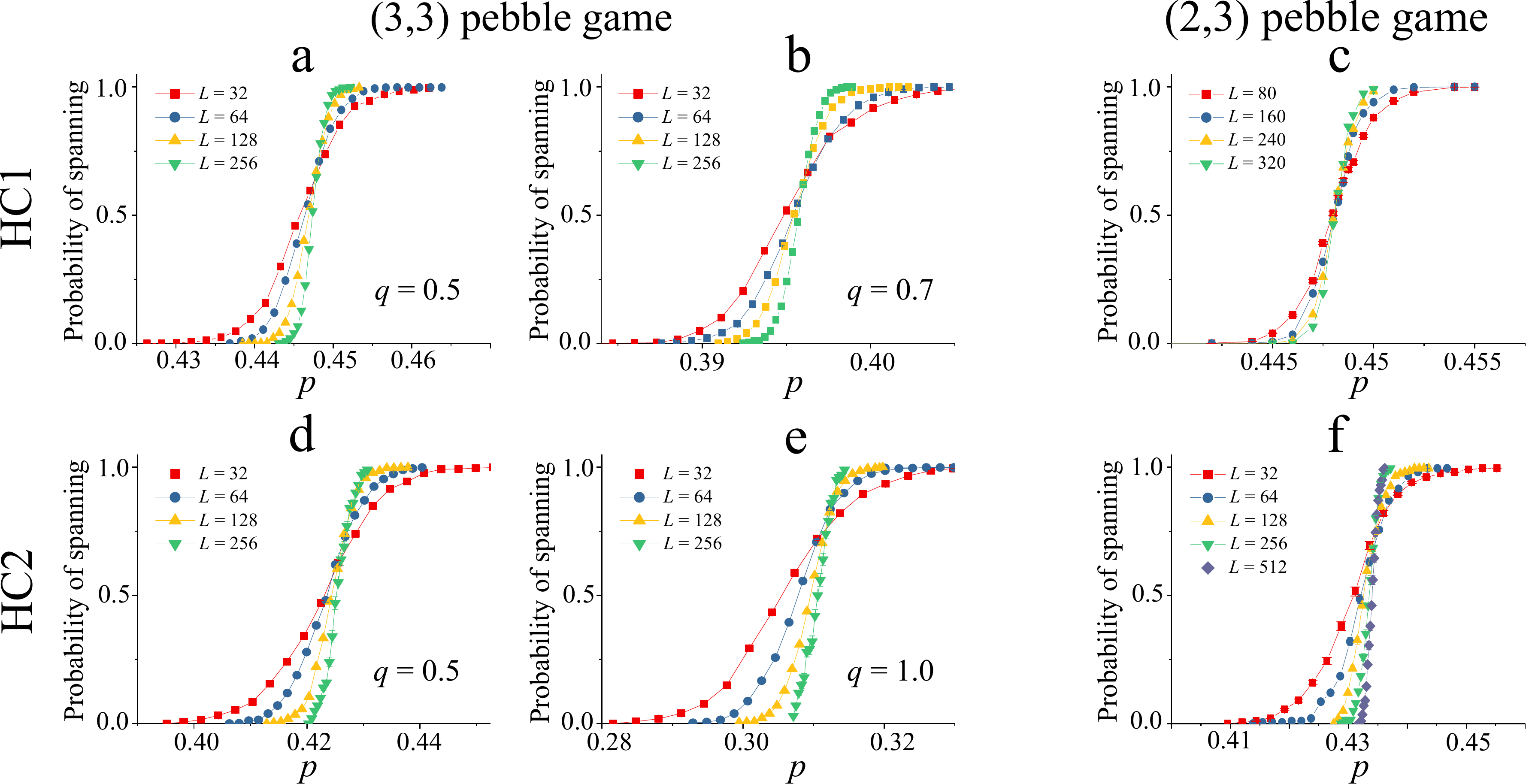}
\caption{Probability of having a spanning rigid cluster as a function of $p$ for lattices of different lengths $L$ for different models, with HC1 in the top row and HC2 in the bottom row and the (3,3) pebble game on the left wile the (2,3) pebble game is on the right. Solid lines are fits to data using error function as a fitting function. Data points are averaged over 2500 samples for the (3,3) game and 1000 samples for the (2,3) game.}
\label{fig:Pspan}
\end{figure*}

Once the frictional and sliding bonds have been identified, we construct a constraint network in which frictional bonds below the Coulomb criterion are denoted as double bonds in the constraint network and sliding bonds at the Coulomb threshold are denoted as single bonds in the constraint network. We then play the $(3,3)$ pebble game on this constraint network in which the first number denotes the number of local degrees of freedom and the second number denotes the number of trivial global degrees of freedom, which does depend on boundary conditions. However, Ref. ~\cite{Henkes2016} found that changing the number of trivial global degrees of freedom from 3 to 2 due to periodic boundary conditions did not signficantly affect the rigid cluster analysis for both frictional and frictionless particles and so we stick with the $(3,3)$ and $(2,3)$ pebble games. 

We illustrate the $(3,3)$ pebble game algorithm using several very simple constraint networks in Fig.~\ref{fig:example33pebble}. A more detailed explanation can be found in Appendix \ref{sec:pebble_game}. Examples or the rigid clusters we find below, at, and above the rigidy transition for both HC1 and HC2 are shown in Figure \ref{fig:rigidclusters2}. To compare frictional rigidity percolation with central force rigidity percolation, we complement our analysis with an approach where any double bond is converted to a single bond and a (2,3) pebble game is played since each site contains now only two degrees of freedom. Examples of the corresponding rigid clusters are shown in Figure \ref{fig:rigidclusters3}.

We now implement the common tools percolation analysis for the spanning rigid cluster to quantify the transition for $HC1$ and $HC2$ for different $q$s for the $(3,3)$ pebble game and for the $(2,3)$ pebble game.

\subsection{Results}

\subsubsection{Spanning probability}
We first identify the location of the rigidity transition 
by determining whether or not there exists at least one spanning rigid cluster in the $x$ or $y$ direction as both $p$ and $q$ are varied.  We do this for all four variants, $HC1$ and $HC2$ for both the frictional $(3,3)$ game and the frictionless $(2,3)$ game. For $HC2$ we study $q=0.5$ and $q=1.0$, the two extreme values of $q$. For $q=1$, isostaticity predicts $p_{c}(1)=\frac{1}{3}$. For the $HC1$ model, this is identical to the initial occupation of the honeycomb lattice backbone, and for this regular lattice, we expect one spanning rigid cluster with a unity probability of spanning, i.e. a first order transition. Therefore, for $HC1$, we study $q=0.5$ and $q=0.7$ and do not explore the limit $q\rightarrow 1$.

\begin{figure*}[t]
\includegraphics[width=0.95\textwidth]{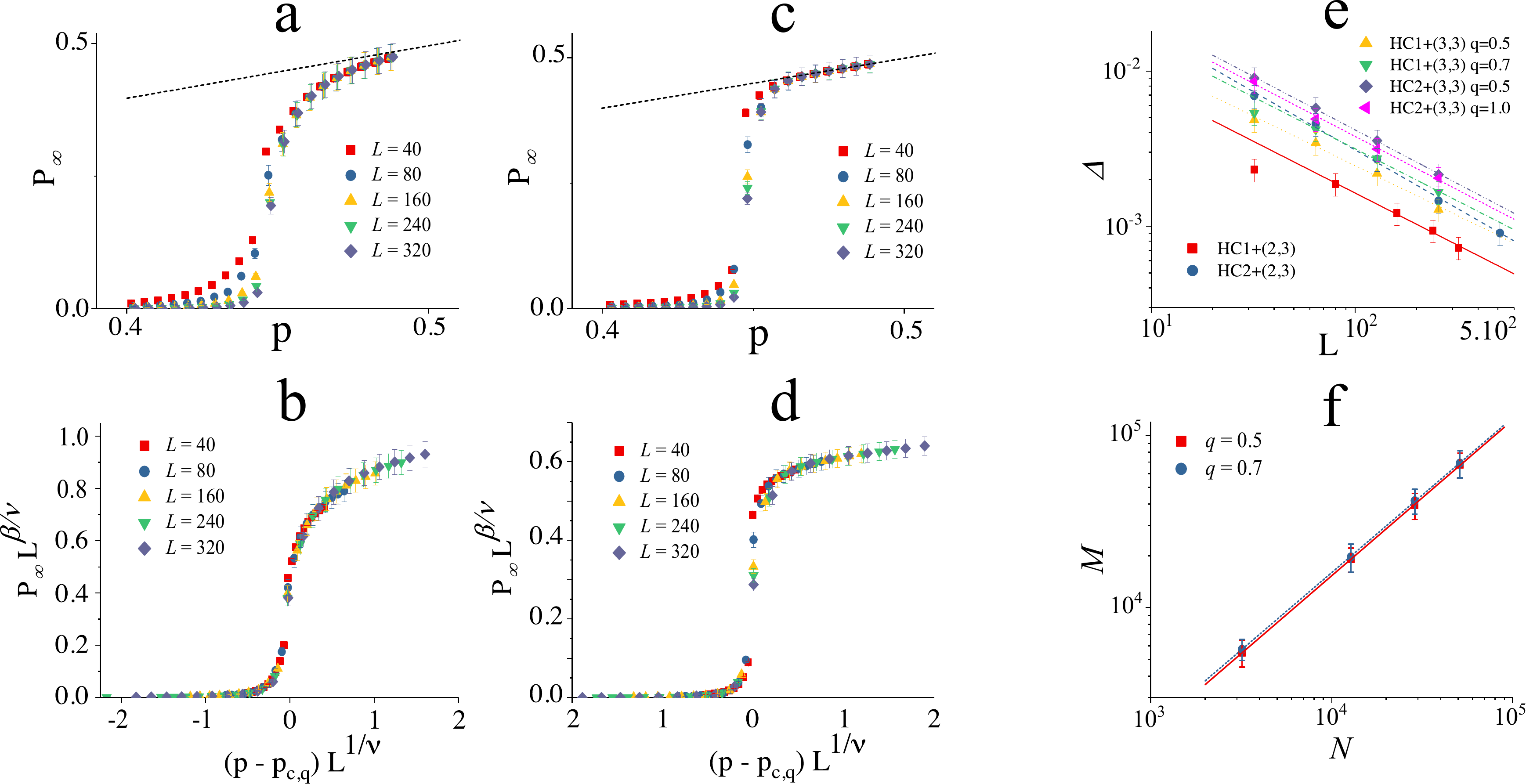}
\caption{(a) For $HC1$ with $q=0.5$, we plot $P_{\infty}$, the fraction of occupied bonds in the largest rigid cluster, as a function of $p$ for different system sizes.  We observe $P_{\infty} \sim (p-p_{c}(q))^{\beta}$ just above critical point and tends towards $p$ further away from the transition. (b) Collapse of (a) using $p_{c}(q)=0.448$, \(\nu =1.56\), \(\beta=0.18\). (c) $P_{\infty}$ for $HC1$ with the $(2,3)$ pebble game as a function of $p$ for different system sizes. (d) Collapse of (c) using $p_{c}(q)=0.448$, \(\nu =1.54\), \(\beta=0.07\). (e) $\Delta$, as defined in eq. \ref{eq:Width_nu}, versus system length $L$ for six different cases of the model. (f) Log-log plot of the number of bonds in the spanning cluster $M$ versus $L$ for $HC1$ with $q=0.5$ and $q=0.7$.}
\label{fig:Lscaling}
\end{figure*}

Figure \ref{fig:Pspan} plots the probability that the system contains at least one spanning rigid cluster as a function of $p$ for different system lengths $L$.  Panel a presents data for $HC1$ with  $q=0.5$, while panel b presents data for $HC1$ with $q=0.7$. In both subfigures, different curves with different system sizes cross near a particular value of $p$, which indicates the location of the transition point denoted hereafter as $p_{c}(q)$. In particular, $p_{c}(0.5)\approx 0.447(1)$ and $p_{c}(0.7)\approx 0.396(1)$.  These two critical points are very close to the results from the generalized isostaticity counting in Equation \ref{eq:pcequation} with $p_{c}(0.5)=\frac{4}{9}\approx 0.444$ and $p_{c}(0.7)=\frac{20}{51}\approx 0.392$. 

The probability of spanning for $HC1$ at $p_c(q)$ for both $q=0.5$ and for $q=0.7$, is approximately $0.6$. Since this value is significantly less than unity, our findings suggest a continuous transition for the onset of the spanning rigid cluster. Typically, the value of probability of spanning at the transition is not a universal quantity and depends on details of the model. For the frictionless version of $HC1$, shown in panel c, the crossing point is more difficult to determine but estimate it to be near $0.448(1)$. 

\begin{figure*}[t]
\centering
\includegraphics[width=0.9\textwidth]{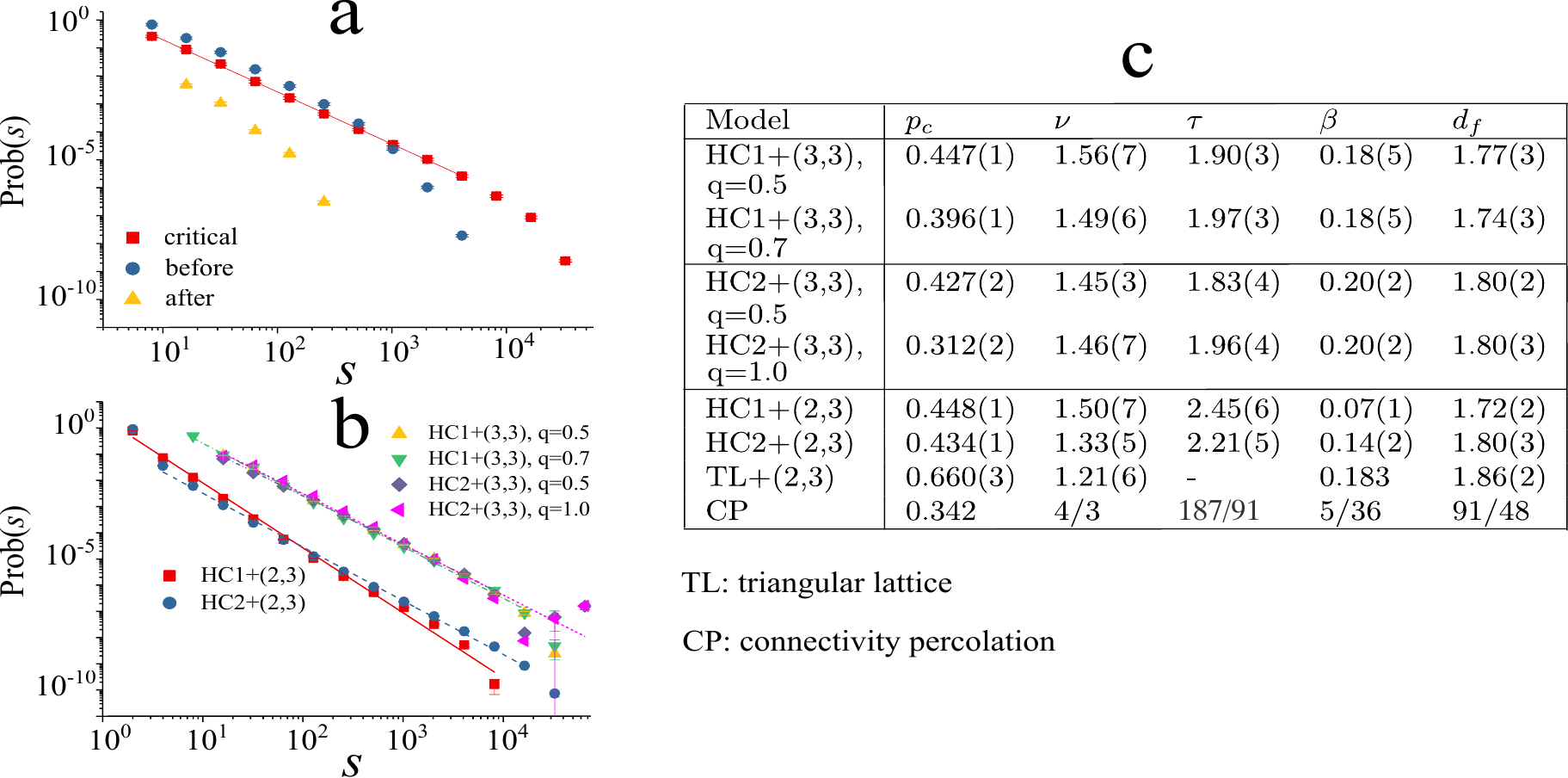}
\caption{(a) $HC1$ with $L=320$ and $q=0.5$, the red squares shows the probability for having finite rigid clusters with different sizes very close to the transition point. Blue circles and yellow triangles show the distribution below and above rigidity transition respectively. (b) Nonspanning cluster size probability distribution at $p_c$ for the six different cases of the model. (c) This table lists the types of rigidity transitions and some critical exponents for the different models defined as the following: $\xi\sim (p-p_{c,q})^{-\nu}$ is the correlation length and diverges at critical point; the nonspanning rigid clusters size may obey a broad distribution, $n_s\sim s^{-\tau}$, at the critical point; $d_f$ is the fractal dimension of spanning rigid cluster at the rigidity transition; $\beta$ is the order parameter exponent; $q$ is the percentage of contacts as double
bonds. For comparison, exponents in rigidity percolation
exponents from the triangular lattice (TL) using the $(2,3)$
pebble game; as well as ordinary
connectivity percolation (CP) exponents are listed here as
a point of reference.}
\label{fig:exponents}
\end{figure*}

For the $HC2$ version of the model (bottom row of Fig.~\ref{fig:Pspan}), we again find crossing points near the predicted generalized isostaticity counting since the formula also applies to this variant of the bond occupation.  Since there is no ordered honeycomb backbone that is initially occupied, we explore both the lower and upper bounds of $q$, i.e. $q=0.5$ and $q=1.0$ (panels d and e). Our results can be found in the first column of the Table in Figure \ref{fig:exponents}. We note that there is greater discrepancy of the estimated $p_c$ from generalized isostaticity for $HC2$ than for $HC1$.  We also note that the probability of spanning at the transition (the crossing point) now differs between $q=0.5$ and $q=1.0$, which does not imply a different universality class because the crossing point depends on details of the lattice.  The frictionless version of $HC2$ is plotted in panel f.

\subsubsection{Correlation length}

The correlation length $\xi$ quantifies how two distant particles/sites interact.  In percolation, it can be extracted from a two-point correlation function for the probability of occupied sites some distance from each other participating in the same rigid cluster.  In a continuous transition, the correlation length diverges at the transition, while near the critical point, $\xi\sim|p-p_{c}(q)|^{-\nu}$ on either side of the transition, where $\nu$ is the correlation length exponent. In a finite-size system and near the transition, $\xi$ is replaced by the system length $L$. For each realization, after this replacement, the system has a finite-size critical point $p_c^L(q)$ when the system contains a spanning rigid cluster, with $|p_c^L(q)-p_c^\infty(q)|\propto L^{-\frac{1}{\nu}}$. Since the location of the transition fluctuates for each realization, we therefore obtain a distribution of finite-size critical points as observed in Figure ~\ref{fig:Pspan}. The standard deviation of this distribution, $\Delta$, yields a measurement of the correlation length exponent~\cite{percolation}. More precisely, 
\begin{equation} 
\label{eq:Width_nu} 
\Delta(L)=\sqrt{\overline{p_{c}^L(q)^2}-\overline{p_{c}^L(q)}^2}\sim L^{-1/\nu}. 
\end{equation}
Using error function fits to the data in Figure \ref{fig:Pspan}, we numerically differentiate the curves and fit to Gaussians to compute $\Delta(L)$ and extract the correlation length exponents $\nu=1.58\pm 0.13$ for $HC1$ with $q=0.5$ and $\nu=1.48\pm 0.20$ for $q=0.7$. Both values are within one standard deviation of each other.  For the frictionless version of $HC1$, we obtain $\nu=1.50\pm0.07$.  For $HC2$, we find $\nu=1.48\pm 0.05$ for $q=0.5$, $\nu=1.43\pm0.04$ for $q=1.0$, and $\nu=1.33\pm 0.05$ for the frictionless version.  Figure \ref{fig:Lscaling}e shows width of the transition for the six different variations of the honeycomb lattice model. 

\subsubsection{Spanning rigid cluster}
We now study the properties of the spanning rigid cluster using $P_\infty$, the fraction of occupied bonds in the spanning rigid cluster. Fig. \ref{fig:Lscaling}a shows $P_{\infty}$ for increasing $p$ for different system lengths for $HC1$ with $q=0.5$. We note that $P_{\infty}$ at $p_c(0.5)$ decreases as the system size increases, which, again, suggests that rigidity transition here is continuous.
The behavior of this curve just above the critical point $p_{c}(q)$ is described by the order parameter exponent $\beta$ with $P_{\infty} \sim (p-p_{c}(q))^{\beta}$ for $p\ge p_{c}(q)$, for an infinite size system. As long as $L>>\xi$, the equation applies and $P_{\infty} \sim \xi^{-\frac{\beta}{\nu}}$. However, when $L<<\xi$, the length scale will be set by $L$ such that $P_{\infty} \sim L^{-\frac{\beta}{\nu}}$. We therefore introduce a universal scaling function $f(\frac{L}{\xi})$ that interpolates between these two regimes, or
\begin{align}
P_{\infty}(p,L,\xi)&=(p-p_{c}(q))^{\beta}f(\frac{L}{\xi})\nonumber\\
&=(p-p_{c}(q))^{\beta}f(L(p-p_{c}(q))^{\nu})\nonumber\\
&=L^{-\frac{\beta}{\nu}}\tilde{f}(L^{1/\nu}(p-p_{c}(q))),
\end{align}
with $f(\frac{L}{\xi})=(\frac{L}{\xi})^{-\frac{\beta}{\nu}}$ for $L<<\xi$ and a constant for $L>>\xi$.  The universal scaling function $\tilde{f}(\frac{L}{\xi})$ can be obtained by rescaling
\begin{equation} 
\label{eq:scaling_func} 
P_{\infty} L^{\frac{\beta}{\nu}}=\tilde{f}((p-p_{c}(q))L^{1/\nu}) 
\end{equation} 
as is done in Fig. \ref{fig:Lscaling}b for $q=0.5$, with $\nu=1.56$ and $\beta=0.18$ used as fitting parameters to obtain the optimal collapse. This estimate for $\nu$ is consistent with our previous measurement for the same $q$ from $\Delta(L)$ in Fig. \ref{fig:Lscaling}e.  The collapse supports the notion of a continuous rigidity transition.  We implement the same protocol for the remaining cases to look for a continuous rigidity transition. 

With the exception of the frictionless version of $HC1$, shown in Fig. \ref{fig:Lscaling}c-d, the order parameter exponent does not vary too much between the different models, as summarised in Fig. \ref{fig:exponents}c, though given the smallness of $\beta$, it is more difficult to measure as precisely as $\nu$.  The small value of $\beta\approx 0.07$ for the frictionless version of $HC1$ perhaps suggests that this model is similar to the square and kagome lattices with next-nearest neighbors studied in Ref. ~\cite{Zhang2015}. There, a hybrid transition was found, where the onset of the spanning cluster was discontinuous, but with a diverging correlation length, though the correlation length exponent appeared to be unity. 

In addition to the order parameter exponent, one can also measure the fractal dimension of the spanning cluster to determine whether or not it is, indeed, fractal. To test for this possibility, the fractal dimension is determined by measuring the number of bonds in the spanning rigid cluster, $M$, as a function of system length such that $M\sim L^{d_f}$. In Figure \ref{fig:Lscaling}f, we see that when $q=0.5$, $d_f=1.81\pm 0.06$. For $q=0.7$, $d_f=1.80\pm 0.05$, so we observe little change in the fractal dimension with $q$, at least for these system sizes, provided $q$ is not close to unity. Similar fractal dimensions were found for both the frictional and frictionles versions of the $HC2$ version of the model and are listed in Table 1. 

\subsubsection{Nonspanning rigid clusters}

In connectivity percolation, one typically investigates the nonspanning cluster size distribution, defined as the number of finite clusters of size $s$ per lattice site/bond, or $n_s$~\cite{percolation}. At the transition, $n_s \sim s^{-\tau}$, where $\tau$ is the cluster size exponent. For connectivity percolation, $\tau > 2$ strictly.  How can we understand this result? The inequality $\tau>2$ in connectivity percolation is a consequence of the following.  We start with $\sum_{s=1}^{\infty}sn_s(p)+P_\infty(p)=p$. Since $P_\infty(p_c)=0$ for connectivity percolation, then $ \sum_{s=1}^{\infty}sn_s(p_c)=p_c$ at the transition. Using the assumption that $n_s \sim s^{-\tau}$ and converting the sum to an integral, $\tau>2$ for convergence to a finite value, i.e. $p_c$.

In rigidity percolation, the situation is more complex because there are non-spanning rigid clusters, spanning rigid cluster(s), and floppy regions.  There are no floppy regions in connectivity percolation. If both the nonspanning rigid cluster size distribution and the floppy cluster size distribution are power laws independently at the transition, each exponent associated with the respective size distribution should be greater than 2.  On the other hand, if one of the exponents is less than 2, that would suggest a natural cutoff for that type of cluster and the more tenuous structure could still facilitate a continuous transition.  If both exponents are less than 2, then perhaps this aspect of the transition is discontinuous. 

Also, when looking at how nonspanning rigid clusters grow/enlarge, the addition of a bond between two finite rigid clusters does not imply a larger rigid cluster formed by simple addition as it does in connectivity percolation, which makes it difficult to study how two nonspanning rigid clusters merge and become one. We illustrate the complexity of such a merging in Figure \ref{fig:merging}c-d where adding one double bond merges five rigid clusters and some floppy regions into a single spanning rigid cluster.  Merging through simple addition also what allows one to derive relations between the nonspanning cluster size distribution exponent and the fractal dimension of the spanning rigid cluster, for example, otherwise known as a hyperscaling relation. Given the nontrivial merging for rigidity percolation, we cannot rely on the hyperscaling relations developed for connectivity percolation. We return to the consequences of this phenomenon below.

Here, we keep track of the nonspanning rigid clusters only and posit that their size distribution also behaves as a power law at the transition with exponent $\tau$, as above.  If $\tau<2$, then there is presumably a characteristic cutoff for the nonspanning rigid cluster sizes at large enough sizes with coupling to the floppy regions perhaps driving the continuity of the transition. Figure \ref{fig:exponents}a shows the probability for having a nonspannning rigid cluster of size $s$ as $p$ is increased through the transition point for $HC1$  with \(q=0.5\) on a log-log scale. Below the transition point, there are many small rigid clusters in the system. As $p$ increases, they merge into larger ones and the distribution broadens to approach a linear function on a log-log scale; the downward trend of the tail is due to finite size effects. We obtain $\tau=1.90\pm 0.03 < 2$ from a linear fit to the relevant part of the curve. Above $p_{c}(q)$, the spanning rigid cluster ``swallows'' the nonspanning rigid clusters and ultimately, as $p$ is increased far beyond the transition point, there is only one spanning rigid cluster left.  We have measured $\tau$ for the six different cases and find a persistent difference between the frictional and frictionless case in that $\tau>2$ for the frictionless cases, while $\tau<2$ for all frictional versions (see Fig. \ref{fig:exponents}c) indicative of rather different ways the rigid clusters merage and grow in the two cases.

\subsubsection{Summary}

The results of our finite-size scaling analysis are summarized in the Table in Fig. \ref{fig:exponents}. We also include the exponents for rigidity percolation using the (2,3) pebble game on the triangular lattice (denoted as $TL$) and for connectivity percolation (denoted as $CP$) on the triangular lattice for comparison.  For the frictional versions, i.e. the (3,3) pebble game, we find that $HC1$ and $HC2$ appear to be in the same universality class, with the exception of the special case of $HC1$ at $q=1$ in which a discontinuous transition emerges as discussed earlier in Sec. II.  We also conclude that exponents associated with $HC2$ and the (2,3) pebble game are in the same universality class as the exponents for rigidity percolation on the triangular lattice obtained about twenty years ago.  On the other hand, we find that the exponents associated with $HC1$ and the (2,3) pebble game are potentially more related to the square lattice plus braces (i.e., next-nearest neighbors) in which a hybrid transition was found~\cite{Zhang2015}, so that this case is special, just as $HC1$ with $q=1$ is special.  

\begin{figure}[htb]\centering
\includegraphics[width=0.85\columnwidth]{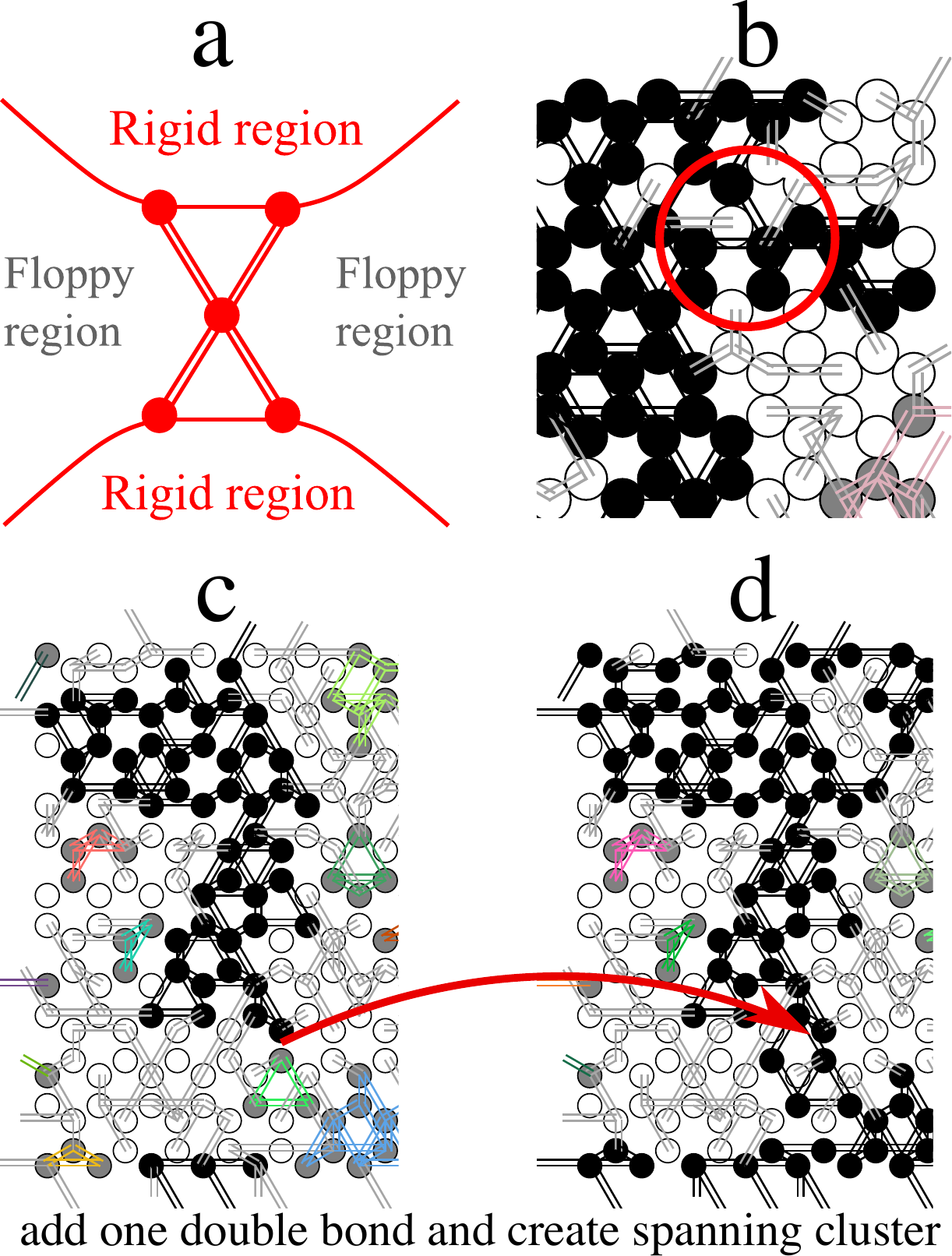}
\caption{Rigid cluster merging mechanism for the (3,3) pebble game. (a) Schematic hinge linking two rigid clusters. (b) Example of a hinge in HC2. (c)-(d) Adding exactly one double bond merges and grows five smaller rigid clusters into a new spanning rigid cluster. }
    \label{fig:merging}
\end{figure}

Focusing now on frictional rigidity percolation versus central-force rigidity percolation, we conclude that they represent two distinct universality classes based on the rather different values of $\nu$ and $\tau$.  The order parameter exponents $\beta$ are not very distinct, which could either be significant, or be due to smallness of the number which makes it difficult to discriminate between models. Setting the order parameter exponent aside for now, let us ask what mechanism could drive frictional rigidity percolation and central-force rigidity percolation to be in distinct universality classes? To begin to answer this question, we ask the following question: How do two rigid clusters combine to form one larger rigid cluster? Unlike in connectivity percolation, for both the (2,3) and (3,3) pebble games, two independently rigid clusters cannot become one rigid cluster by joining via a single bond. For a (3,3) pebble example with $q=1$, consider two triangles, which are individually rigid. If they are now joined by a double bond, 18 degrees of freedom of the now 6 particles, minus 3 global degrees of freedom, are compared with 14 constraints from 7 bonds to ultimately give one floppy mode.  However, for the (3,3) pebble game two independent rigid clusters connected by two double bonds makes a new rigid cluster, which is not the case for the (2,3) pebble game. Rigid hinges are another means by which two rigid clusters can merge at a point and still be rigid. In the (2,3) pebble game version, hinges consisting of single bonds between rigid clusters are always floppy, and so rigid hinges cannot exist. However, in the (3,3) pebble game, this is not true, at least for a hinge comprised of all double bonds, see Fig.~\ref{fig:merging}a-b.  

We conclude that two rigid double bonds and the rigid hinge (composed of double bonds) are distinct means of propagating rigidity in the (3,3) pebble game (frictional rigidity percolation) that do not occur in the (2,3) pebble game. Both motifs are more localized than possible propagators in the (2,3) pebble game and allow for rigid clusters to merge at a point or along a line, even in the absence of floppy regions.  The presence of floppy regions complicates matters, see the example in Fig.~\ref{fig:merging}c-d.  Here, the addition of one double bond brings transmits rigidity across five nonspanning rigid clusters with connected via floppy bonds prior to the additional one double bond. Given these local motifs of connecting rigid clusters, one can understand why the sizes of nonspanning rigid clusters are typically larger for the frictional rigidity percolation as compared to central-force rigidity percolation. 

And yet, even though the rigid cluster formation differs between the frictional and central-force rigidity percolation, can the differences in the order parameter exponent be that minimal?  
In addition to observing little variation in $\beta$ for the different models studied, we also observe little variation in $d_f$ for both HC1 and HC2 studied with just central-forces and with frictional forces. Typically, for continuous transitions, there exist hyperscaling relations between the different exponents so that if $\tau$ is different, then one would expect $d_f$ to be different and, therefore, $\beta$ to be different as well. Such relations are rigorous for connectivity percolation. For rigidity percolation, they have yet to even be derived since the problem is considerably more complex. For example, to derive the hyperscaling relation between $\tau$ and $d_f$ in connectivity percolation,  there are only two types of connected clusters---those that are finite (nonspanning) and those that are spanning.  In rigidity percolation, there are however three types of entitites with the floppy regions playing a role in determining how two different rigid clusters join up to form one larger rigid cluster.  Moreover, for $\tau<2$, we expect a characteristic size cut-off. Without such hyperscaling relations, it is difficult to evaluate the significance of our findings for $\beta$ and $d_f$.

\begin{figure}[htb]\centering
\includegraphics[width=0.99\columnwidth]{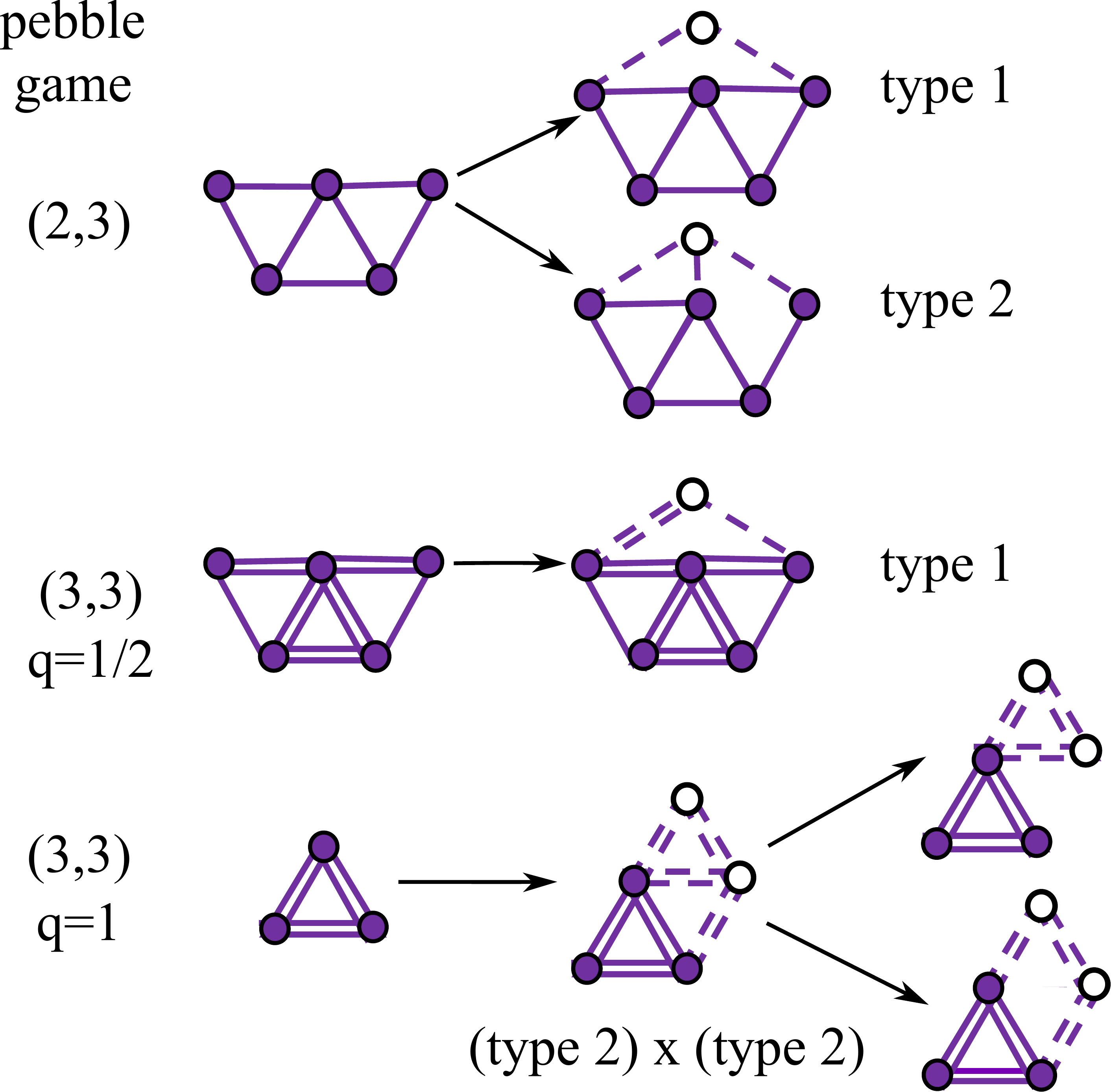}
\caption{Schematic of Type I and Type II Henneberg moves for the (2,3) game (top) and for the (3,3) game (middle and bottom).}
\label{fig:Henneberg}
\end{figure}

\section{Minimal rigidity proliferation}

As mentioned above, the order parameter exponents and the fractal dimension are not very distinct between frictional and central force rigidity percolation. Are they in fact the same, signalling features of superuniversality for this aspect of the transition? Or is it the case that the order parameter exponents are distinct but the distinction is small, making it hard to detect?  If we can find a limiting case where the order parameter exponents are actually the same, this strengthens the case for superuniversality rather than relying purely on numerical analysis. So let us now explore more explicitly connections between frictional rigidity percolation and central-force rigidity percolation via a subset of configurations using an algorithmic approach rather different from finite-size scaling.  As will become clear below, connectivity percolation also enters the picture, since if we can construct spanning rigid clusters in the same way as geometrically connecting clusters, then we have evidence for superuniversality across all three models.

To find out if this approach is feasible, we need to construct our new algorithm step by step. Let us first review invasion percolation, which is motivated by the problem of one fluid displacing another from a random, porous medium~\cite{Wilkinson1983}. More importantly for us, invasion percolation allows one to create a spanning cluster on a lattice that has the same properties as a spanning cluster in connectivity percolation. Next, we will review the Henneberg moves~\cite{Henneberg1911}, which are used to grow a large minimally rigid network (a Laman graph) from a small minimally rigid network in the central force case. We will then extend the Henneburg moves to include frictional forces and ultimately unify the two concepts, invasion percolation and Henneburg moves. The final algorithm that we introduce, minimal rigidity proliferation (MRP), allows us to grow minimally rigid networks that span a frictional system, and \emph{only} grow such networks.

\begin{figure*}[t]
\centering
\includegraphics[width=0.99\textwidth]{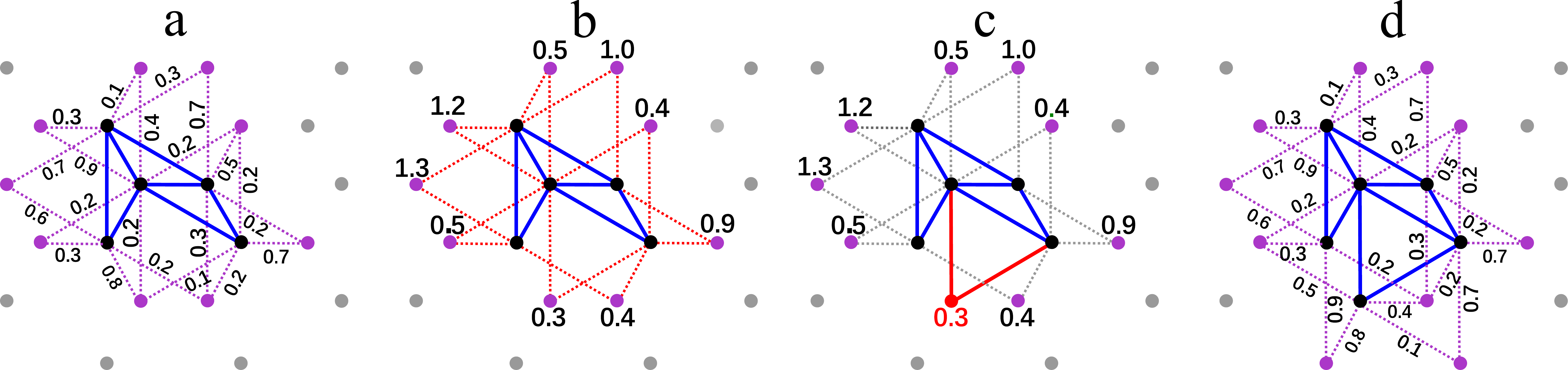}
\caption{Schematic of minimal rigidity proliferation (MRP). (a) Existing rigid cluster (blue) surrounded by nearest and next-nearest bonds with their respective weights (purple). (b) Minimal sum of weights from bond pair has been associated to sites; candidate Henneberg move pairs are in red. (c) The move is executed at the site with the lowest total weight. (d) New rigid cluster surrounded by nearest and next-nearest bonds. }
\label{fig:imrp}
\end{figure*}

Invasion percolation is a modified version of connectivity percolation where the spanning cluster grows along the path of smallest weights, with the following algorithm:
\begin{enumerate}
\item Assign uniformly distributed random numbers ranging from $0$ to $1$ to bonds on a lattice as their weights.
\item Occupy an initial bond, and create a list of all its neighbors. This list creates a boundary of bonds. 
\item Occupy the bond from the list that has the smallest weight. 
\item Update the list so that it contains all unoccupied nearest neighbors of occupied bond.
\item Repeat 3 and 4, until the occupied cluster spans the entire lattice. 
\end{enumerate}

The above algorithm reduces to the Leath algorithm~\cite{Leath1976}, which creates the spanning cluster for bond connectivity percolation for $p>p_c$ in the following limit: Instead of occupying the boundary bond with the smallest weight, all boundary bonds whose weight is less than $p$ are accepted into the cluster, and then the boundary list is updated. The algorithm terminates when there are no bonds on the boundary with weights less than $p$.
This modification from invasion percolation to the Leath algorithm does not affect the large scale structure of the spanning cluster, i.e. they remain part of the same universality class~\cite{Wilkinson1984}.

Let us also review the Henneberg moves for building a minimally rigid network associated with frictionless particles, i.e. for central forces only. A minimally rigid graph in this case is also known as a Laman graph. Minimal rigidity occurs when the degrees of freedom match the constraints and there are no rendundant bonds, as determined through a $(2,3)$ pebble game. Starting from such a network $G(N,N_B)$ with $N_B$ bonds and $N$ sites, one can extend it using two basic Henneberg moves as illustrated in Figure \ref{fig:Henneberg} (top):

\begin{itemize}
\item add one site and two bonds between this site and two points in $G$ , then $G'(N+1,N_B+2)$ is the new minimally rigid network (Type I move).
\item or add one site and three bonds between this site and three prior sites in $G$, then delete a prior bond between two of the selected three prior sites (Type II move).
\end{itemize}
Both moves simultaneously add two degrees of freedom and two constraints, which results in a minimally rigid graph by induction.

Now we generalize, for the first time, Henneberg moves for the (3,3) pebble game in order to propagate minimal rigidity. We focus on two cases: $q=1/2$ and $q=1$. For $q=1/2$, we consider only a Type I move by adding a site and then adding three bonds, one double bond and one single bond (see Figure \ref{fig:Henneberg}, middle).  This move perpetuates minimal rigidity since no dependent constraints are introduced. For $q=1$, i.e. all double bonds, we consider two Type II moves in series, if you will, by adding two sites, where the first site connects to two existing sites and the second new site must attach to the initial new site as well as an older site. Then, any one of the double bonds between the first new site and either old site is removed, though not the bond between the two new sites, to preserve minimal rigidity (see Figure \ref{fig:Henneberg}, bottom).

Having discussed invasion bond percolation and the ``growing'' of minimal rigidity via generalized Henneberg moves, we are now ready to introduce \emph{minimal rigidity proliferation}.  We will first focus on $(2,3)$ minimal rigidity and then address $(3,3)$ minimal rigidity. To create a spanning minimally rigid cluster as defined by the $(2,3)$ pebble game, we combine the Henneberg move Type I and invasion bond percolation in the following algorithm (see Fig. \ref{fig:imrp} for an illustration):

\begin{enumerate}
\item Assign uniformly distributed random numbers ranging from $0$ to $1$ to bonds on the honeycomb lattice as their weights.
\item Begin by occupying a random triangle between three closest sites and create a list of all nearest and next-nearest neighbor bonds of these sites.
\item Determine the sum of the weights of any two bonds from the sites on the list that join at one site and the existing sites in the graph, find the smallest sum and occupy those two bonds.
\item Update the bond list such that it contains any unlisted nearest and next-nearest neighbor bonds of the newly added site.
\item Repeat 3 and 4, until the graph spans the lattice.
\end{enumerate}

\begin{figure}[htb]\centering
\includegraphics[width=0.99\columnwidth]{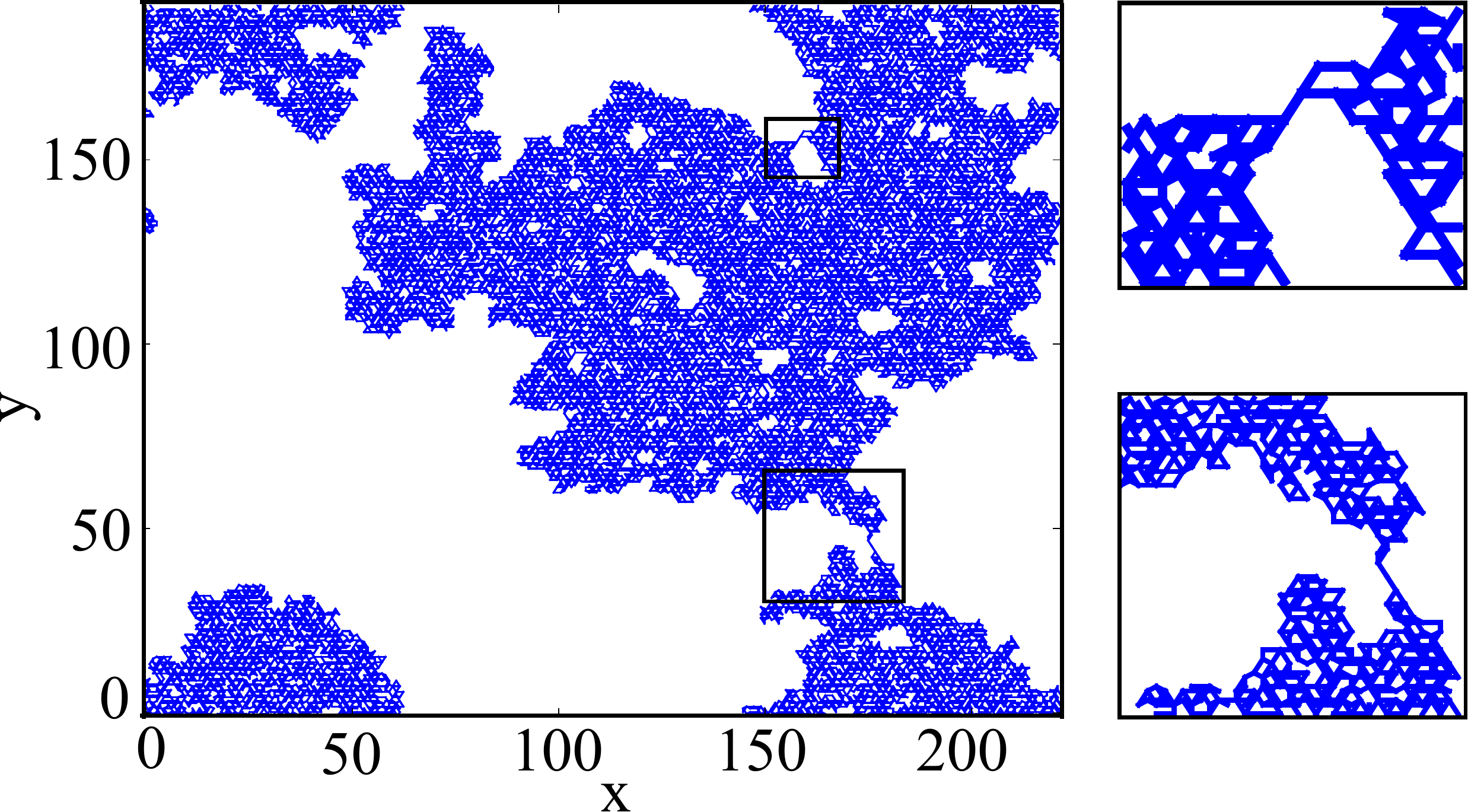}
\hfill
\caption{An example of spanning rigid cluster constructed using minimal rigidity proliferation; two examples of rigid hinges are shown in more detail on the right.}
\label{fig:ExampleIRP}
\end{figure}

\begin{figure}[htb]\centering
\includegraphics[width=0.99\columnwidth]{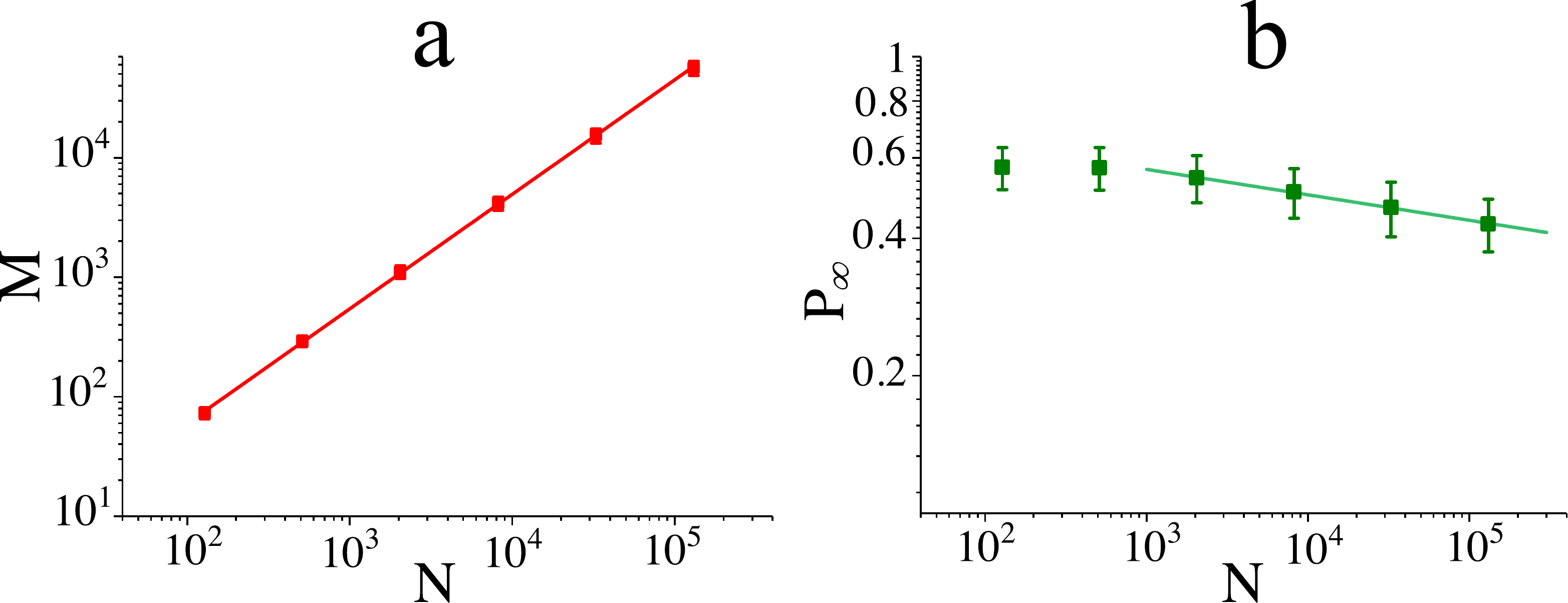}
\caption{(a) Log-log plot between size of spanning cluster at the critical point $M$ and size of whole system $N$. The slope of less than one indicates that the spanning cluster at the critical point has a fractal dimension $d_f=1.916\pm 0.010$. (b) As the system size increases, $P_{\infty}$ goes to zero, suggesting a continuous transition. All plots have been averaged from $935$ samples.}
\label{fig:IRP3}
\end{figure}

Though the graph is grown by adding two bonds at a time, as opposed to one, we still expect that this process will fall under the connectivity percolation universality class. Why? Because adding two bonds (with their additive weights) at a time involves a rescaling of time in which two bonds are added in one time step as opposed to two bonds in two time steps. Moreover, such a local modification from invasion percolation shouId not alter the long wavelength behavior which captures the universality class. In Fig. \ref{fig:ExampleIRP} we show an example of a spanning minimally rigid cluster on the honeycomb lattice with NNN bonds using minimal rigidity proliferation. We also measure the fractal dimension by computing the number of sites in the cluster $M$ as a function of total number of sites $N$, as shown in Figure \ref{fig:IRP3}a, and obtain  $M=N^{0.958}$. Since $N=L^2$, this leads to a fractal dimension of the spanning rigid cluster at the critical point of $d_f=1.916$. Figure \ref{fig:IRP3}b shows $P_{\infty}$, the fraction of the system in the spanning cluster, converging to zero when system becomes infinitely large, suggesting a continuous transition.

Networks constructed in this way correspond to the frictionless case. Since we have extended the Henneberg moves to the $(3,3)$ pebble game for $q=1/2$ and $q=1$, we can generalize minimal rigidity proliferation to the frictional case. For the $q=1$ case, two Henneberg Type II moves are made in sequence to arrive at one gowth step.  With bond removal, it is not immediately clear that the minimally rigid cluster growth results in the same cluster structure as the $q=1/2$ case, and so we leave this for future study. However, since the Type I move for $q=1/2$ corresponds precisely to the Type I move for the central force case, we expect the same configurations as above, just with half single and half double bonds. Both central-force percolation and frictional rigidity percolation collapse to an \emph{identical} construction in this case.  We have already argued that adding two bonds at a time involves a rescaling of time from adding one bond at a time for central-force percolation, so that we also expect the $q=1/2$ frictional process to be in the same universality class as connectivity percolation. In other words, within this subset of minimally rigid configurations, we expect superuniversality to emerge: All three universality classes collapse into one!  Interestingly, transfer matrix methods (not focusing on minimally rigid clusters) argued that connectivity and central-force percolation were in the same universality class but their results were later discounted~\cite{Roux1988,Jacobs1995}.  We now have some understanding as to why some exponents appear to be quite close in value. 

\section{Hierarchical lattices}

While we have presented predominantly numerical results so far, one exactly solvable rigidity percolation model is rigidity percolation on hierarchical lattices. However, such lattices have the particular property that the exponents depend on details of the lattice. To determine if this property prevails in the frictional $(3,3)$ pebble game, we will first review prior results using the $(2,3)$ pebble game with central forces only and then generalize. 

\begin{figure*}[t]
\centering
\includegraphics[width=0.99\textwidth]{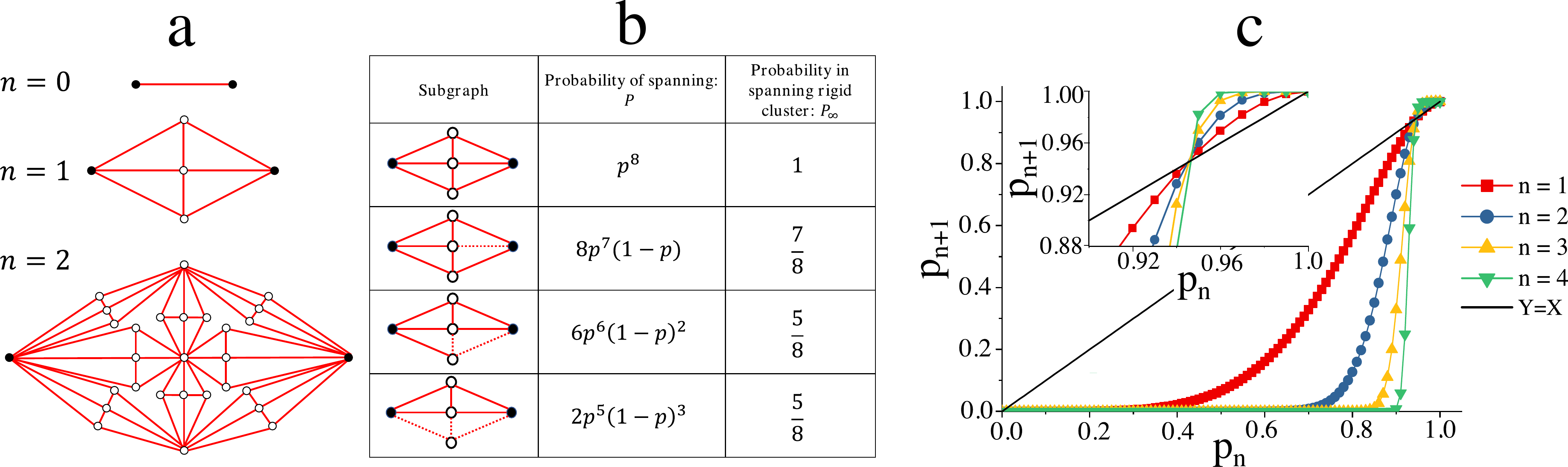}
\caption{(a) First three generations of hierarchical Berker lattice. (b) Subnetwork counting: dashed bonds are not occupied. Every type of subnetwork is a way to obtain a spanning rigid network between two ends(black dots) and its probability is calculated, as well as the probability for an occupied bond to be in the spanning rigid cluster. (c) First four $p_n$ as function of $p$. $p_n$ tends to converge to a step function at $p_c=0.9446$ which jumps from $0$ to $1$. }
\label{fig:hierarchical_23}
\end{figure*}

\subsection{Review: central forces only}
It has been previously shown that central force rigidity percolation transitions in such lattices are a continuous transition~\cite{Barre2009,Stinchcombe2014}. To understand this finding, let us start with the generation of a particular hierarchical lattice known as the Berker lattice~\cite{Stinchcombe2014}. Given two points and a bond as in Figure \ref{fig:hierarchical_23}a, replace the bond with some base structure to generate a first generation hierarchical structure. This replacement continues ad infinitum to arrive at a network with an infinite number of sites between two initial points. For this particular lattice, the $n^{th}$ generation contains $8^n$ bonds (with the exception of $n = 0$).  To embed this lattice in two-dimensions, the bond length decreases with each generation. 

To analyze rigidity in this hierarchical lattice, assume each bond has a probability $p < 1$ to be occupied. In the $n = 0$ graph, the probability of having a spanning rigid cluster between the two ends (black dots) is $p_0 = p$. In the $n = 1$ graph, the probability of being rigid between two ends can be found by subgraph counting: If all eight bonds are occupied in the $n = 1$ network, there is a spanning rigid cluster between the two ends. The probability of such a structure existing is $p^8$, while the probability for any bond belonging to the spanning rigid cluster is $1$. All other subgraphs that contain a spanning rigid cluster, as determined by the $(2,3)$ pebble game, and their respective probabilities are listed in Figure \ref{fig:hierarchical_23}b. Summing up all ways of having a spanning rigid cluster between the two ends of the $n = 1$ graph, we obtain
\begin{equation*}
p_1=2p^5+2p^7-3p^8.
\end{equation*}

Given the hierarchical structure of the lattice, it is trivial to generalize this relation to 
\begin{equation}
\label{eq:recur_p} 
p_{n+1}=2p_n^5+2p_n^7-3p_n^8, 
\end{equation} from which we can solve for a fixed point, $p_c=0.9446$, 
as the system approaches the thermodynamic limit, i.e. $p_{n+1}=p_n$. 

In Fig. \ref{fig:hierarchical_23}c, $p_n$ as a function of $p_{n−1}$ is plotted for the first four generations. We observe that the curves cross at $p = p_c$ and $p_n$ will converge to a step function which jumps from $0$ to $1$ at $p_c$ as $n$ goes to infinity. Meanwhile, we use $P_R(p)$ to denote the probability for a bond to belong to the spanning rigid cluster. The recurrence relation for $P_R(p)$ is
\begin{equation} 
\label{eq:recur_P} 
P_{R,n+1}(p)=\frac{1}{4}\left(5p_n^4+13p_n^6-14p_n^7\right)P_{R,n}=\lambda P_{R,n}(p), \end{equation}
and near $p_c$, $\lambda=0.9554<1$ demonstrating that the probability of a bond belonging to the spanning rigid network will approach zero as $p$ approaches $p_c$. This trend suggests a continuous transition.

Expanding about $p_c$ in both Eqns. \ref{eq:recur_p} and \ref{eq:recur_P} leads to $(p_{n+1}- p_c) = \lambda_1(p_n-p_c)$ and $P_{n+1}(p) = \lambda_2 P_n(p)$ such that $\lambda_1 = b^{1/\nu}$, $\lambda_2 = b^{-\beta/\nu}$,  and $\lambda_3=b^{d_f}$, where $b$ is the length rescaling factor from one generation of the hierarchical lattice to the next. For the Berker lattice, $\lambda_3=8$. We can therefore determine $\beta=-log(\lambda_2)/log(\lambda_1)$ and $\nu d_f=log(\lambda_3)/log(\lambda_1)$ , which are both quantities that are independent of $b$, resulting in $\beta=0.078$ and $\nu d_f =3.533$. 

\subsection{Frictional forces}

\begin{figure}[htb]\centering
\includegraphics[width=0.99\columnwidth]{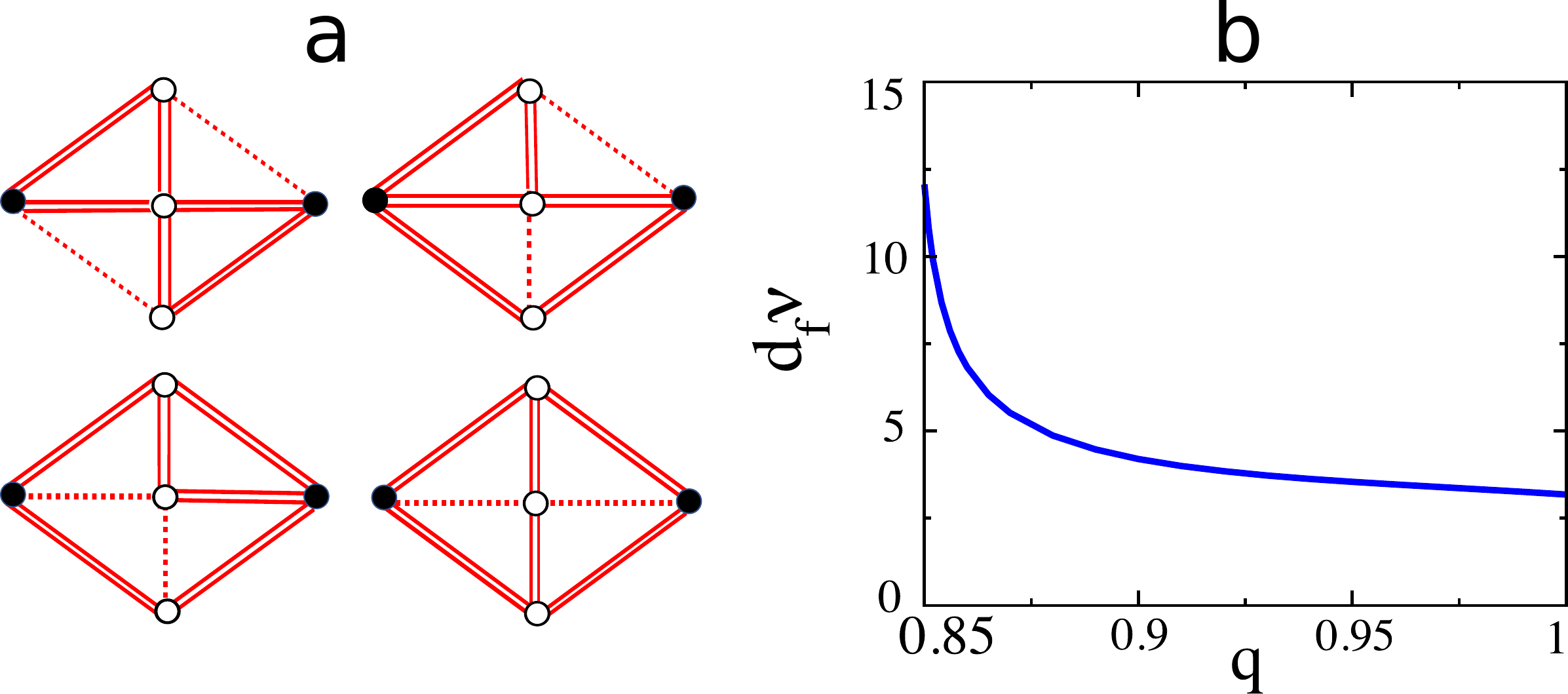}
\caption{(a) Allowed rigid subgraphs for the $q=1$ case with the $(3,3)$ pebble game that are not allowed with the $(2,3)$ pebble game. (b) Plot of $d_f\nu$ versus $q$ for frictional rigidity percolation on the Berker hierarchical lattice.}
\label{fig:bowtie}
\end{figure}

Let us now consider a ``frictional'' hierarchical lattice with double and single bonds to denote frictional and sliding contacts. Double bonds are introduced at random with probability $q$. When double bonds are taken into account, they affect subgraph constraint counting in the hierarchical lattice as we now play the $(3,3)$ pebble game to determine whether or not a subgraph has a spanning rigid cluster. Since there is an increased number of possible subnetworks in the frictional case given that the occupied bonds can be either double or single bonds, let us first discuss the $q = 1$ case. Here, there are several additional type of subgraphs containing a spanning rigid cluster that were not allowed in the central force case, as shown in Fig. \ref{fig:bowtie}a. For instance, one of the subgraphs is not allowed in the central force case because it contains a hinge structure. In the frictional case, the frustrated loops of odd numbers of vertices, or ``gears'', prevent rotation. These additional rigid subgraphs contribute an additional $16p^6(1 − p)^2$ to the probability of having a spanning rigid cluster above the central force case. Therefore, the counting for $q=1$ leads to the recursion relation
\begin{equation}
p_{n+1}=13p_n^8-30p_n^7+16p_n^6+2p_n^5.
\end{equation}
In the limit $n\rightarrow \infty$, we find the unstable fixed point $p_c(q\!=\!1)=0.8533$, in addition to two stable fixed points at $p=0$ and $p=1$.   Moreover, we can compute $P_{R,n+1}(p,q\!=\!1)$ to arrive at
\begin{align}
P_{R,n+1}&=\frac{1}{4}(5p_n^4+48p_n^5-83p_n^6+34p_n^7)P_{R,n}\\
&=\lambda_2 P_{R,n}, \nonumber
\end{align}
such that $\lambda_2(q=1)=0.3511$.  Since $\lambda_2(q=1)<1$, the rigidity transition is continuous with $d_f\nu=3.181$. 

Now we consider $q<1$.  After keeping track of what subgraphs are rigid between the two black circles in Figs. \ref{fig:hierarchical_23}b and ~\ref{fig:bowtie}a in the presence of both double and single bonds, we obtain
\begin{align}
 p_{n+1} &= p_n^8(35q^8-244q^7+474q^6-312q^5+60q^4) \nonumber \\
&+p_n^7(84q^7-210q^6+96q^5) +16p_n^6q^6\label{eq:recur_p_33} \\
&+p_n^5(-8q^5+10q^4). \nonumber
\end{align}

With $q=1$, the unstable fixed point occurs at $p=p_c(q=1)=0.8533$ with the two stable fixed points at $p=0$ and $p=1$. Therefore, $p_n$ will converge to a step function as $n\rightarrow\infty$. However, for $0.8465<q<1$, both the unstable and nonzero stable fixed points, $p_{lower}$ and $p_{higher}$ respectively, are smaller than $1$ so that $p_n$ will converge to a step function which jumps at $p_{lower}$ from $0$ to $p_{higher}$. The reason that $p_{higher}$ is not unity in these cases is because $p$ denotes a double or a single bond such that when $q=1$, $p=1$ translates to all double bonds; however, when $q<1$ and $p=1$, $p_{higher}$ depends on the ratio of double to single bonds. When $q=0.8465$, $p_{lower}=p_{higher}$, which means when $q\leq 0.8465$, $p_n$ will always converge to zero and rigidity transition will vanish entirely, showing that the existence of a rigidity transition in this hierarchical lattice very much depends on $q$. We also compute $d_f\nu$ (for $q>0.8465$) and find that its value depends on $q$, as shown in Fig. \ref{fig:bowtie}b. 
The fact that the correlation exponent depends continuously on $q$ is not necessarily unique as has been found in Ising models on hierarchical lattices, for example~\cite{Kaufman1981}. This sensitivity is presumably due to the special nature of the hierarchical lattice, as detailed in Appendix \ref{sec:general_hierarchical}.

\section{Discussion}

We have now expanded the notion of rigidity percolation to include friction in two dimensions with the extension of the $(2,3)$ pebble game to the $(3,3)$ pebble game and the incorporation of double bonds, which are representative of contacts that are below the Coulomb threshold. In doing so, we have uncovered a new universality class in the realm of rigidity percolation, namely that of frictional rigidity percolation, which is to be directly compared with central-force rigidity percolation on the same lattice. The apparent sensitivity of the nature of the rigidity transition to the type of lattice and type of force has so far made such a direct comparison between two universality classes not possible. For instance, a direct comparison of central-force rigidity percolation to bond-bending rigidity percolation in terms of the order parameter exponent, etc., has not been done because the pebble game has not yet been applied to bond-bending forces. By expanding the scope of rigidity percolation, the direct comparison presented here should help to formulate a more general framework for rigidity transitions. 

More specifically, we make a direct comparison between central force rigidity percolation and frictional rigidity percolation on honeycomb lattices with additional next-nearest bonds. We find different correlation length and rigid cluster size distribution exponents between the two cases but a statistically similar order parameter exponent but could ultimately be demonstrated to be distinct with larger system size studies. Given the different correlation length and rigid cluster size distribution exponents, we propose that local motifs, such as two double bonds and a rigid hinge composed of double bonds, are ways to connect rigid clusters in frictional rigidity percolation that are distinct from central-force rigidity percolation. The rigid hinge is a zero-dimensional contact between two-dimensional rigid clusters, while the double bond is a one-dimensional contact.   In central-force rigidity percolation, there must be additional supporting bonds to connect up rigid clusters. The lower-dimensional rigid cluster connection mechanisms in frictional rigidity percolation (as compared to central-force rigidity percolation) perhaps drive the distiction between universality classes. For the hierarchical lattice, not only are there two different universality classes for each respective model, the exponents depend continuously on the fraction of the double bonds in the frictional case. 

It is interesting to note that in three-dimensions that central-force percolation is thought to exhibit a first-order rigidity transition~\cite{Chubynsky2007}, though Laman's theorem (as implemented via the pebble game) is not rigorous in three-dimensions. Perhaps the mechanisms for connecting up rigid clusters by small numbers of additional supporting bonds are few and far between, thereby demanding a first-order transition, or one needs to simulate much larger systems to observe any fractal nature of the rigid spanning cluster.  In addition to higher dimensions, one could also ask about how rigid clusters connect up in the general $(k,l)$ pebble game with and without double bonds. Presumably other mechanisms for connecting rigid clusters emerge to warrant additional universality classes not yet discovered. Or, possibly, all $k>3$ collapse to frictional rigidity percolation, at least in two-dimensions. 

Since the difference between the order parameter exponent is small between central-force and frictional rigidity percolation, we combined Henneberg moves (extended here to the $(3,3)$ pebble game) and invasion percolation to construct another new model, \emph{minimal rigidity proliferation} (MRP),  that can be implemented for both central-force and frictional rigidity percolation. The rigid cluster in MRP grows in an additive fashion, unlike rigidity percolation where rigid clusters surrounded by floppy regions may not always connect up in a way that leads to larger rigid clusters. With minimal rigidity proliferation, there are no redundant bonds---the spanning rigid cluster is built in a ``clever'' way, which is to be contrasted with the tuning by pruning approaches~\cite{TuningByPruning}, where springs are removed in a way that results in the least change in the bulk modulus, for example, and the jamming graph approach~\cite{JammingGraph} where minimally rigid clusters with the geometric constraint of local mechanical stability are generated. In minimal rigidity proliferation, the order parameter exponent is the same across connectivity, central-force rigidity percolation, and frictional rigidity percolation. This would be the first time superuniversality is observed in rigidity percolation in a way that goes beyond transfer matrix methods. Our work also suggests that looking at minimally rigid configurations---a subset of all possible configurations within rigidity percolation---represents a paradigm shift in the way phase transitions are viewed in the sense that nested within two distinct universality classes there could be an underlying superuniversality establishing deeper connections between the classes than previously thought. 

Since frictional rigidity percolation was devised to explore the nature of the jamming transition in frictional particle packings, this work compels us to make a rather strong claim that the rigidity transition in frictionless particle packings with purely repulsive central forces is of a different nature than the rigidity transition in frictional particle packings.  In fact, the frictionless case with purely repulsive central forces may indeed be a very special case because even rigid cluster analysis of particle packings with both attractive and repulsive central forces indicate a continuous transition~\cite{Koeze2018}. In frictionless packings, there are no redundant bonds, which makes the constraint counting rather straightforward.  However in frictional packings, the redundant bonds emerge such that the constraint counting is more intricate and, therefore, perhaps more non-mean-field.  It will be interesting to apply our constraint counting algorithm to experimental frictional particle packings to test the applicability of our approach as well as to compare the rigid clusters with dynamical matrix calculations, for example. 

Finally, we are currently exploring a limitation of the $(3,3)$ pebble game~\cite{Lester}.  Specifically, if there are four particles forming a square and all four contacts are below Coulomb threshold, then we have a square with all double bonds (like the middle image in Figure \ref{fig:example33pebble} without the diagonal bond). From the $(3,3)$ pebble game this configuration is floppy with one floppy mode that is a pure spin mode and the particles spinning as gears. However these four particles are rigid under strain, since the pure spin mode does not couple to translations, and so one can play a $(3,4)$ pebble if one is interested only in translational rigidity. More generally, odd loops of double bonds do not contain this pure spin mode, while even loops of double bonds do.  This complication can be remedied by keeping track of even and odd loops of double bonds. Any odd loop intersecting an even loop drives the even loop to become odd, if you will. If there are no even loops after looking at intersections of even and odd loops, then the original $(3,3)$ pebble game is robust at all lengthscales.  Near the transition where system-spanning lengthscales dominate, the initial version of the $(3,3)$ pebble game is also robust as long as there is no cluster-spanning set of even loops, which is unlikely due to the intersection with odd loops such as triangles. Note also that the low-energy normal modes of rigid frictional packings show a rough equipartition between rotational and translational degrees of freedom and do not contain any purely rotational modes~\cite{n_m2}.

In closing, our work opens up many new avenues for exploration in rigidity percolation with new constraint counting methods and the discovery of potentially new universality classes. It also invites us to explore not only rigid regions but floppy regions as well, which may be the key in constructing field theories of continuous rigidity percolation transitions. Finally, our new optimal rigid cluster growth algorithms do not waste material and, therefore, perhaps have a chance of being realized in living matter as well as provide mechanical examples for decision-based cluster growth that may draw links with explosive percolation~\cite{DSouza2015}.

The authors would like to thank D. Lester for useful discussion and D. M. Sussman for helpful comments on an earlier draft.  JMS would like to acknowledge finanical support from NSF-DMR-CMMT-1507938 and SH would like to acknowledge support from BBSRC grant BB/N009150/1.

\appendix
\section{The pebble game \label{sec:pebble_game}}

The constraint counting in Section II assumes that every bond/constraint is an independent one. However, not every bond is an independent constraint in a random network. There may exist some redundant bonds. In order to more accurately locate the critical point where rigidity percolation occurs in two-dimensional networks by keeping track of independent and redundant constraints, one can invoke the pebble game. This algorithm was described in Ref.~\cite{Jacobs1997} and is rooted in the following Laman condition: A two-dimensional network with $N$ sites is minimally rigid if and only if it has $2N-3$ bonds and no subnetwork of $k$ sites has more than $2k-3$ bonds~\cite{Laman1979}. To implement the Laman condition numerically requires checking all possible subnetworks, which is comptuationally expensive.  The pebble game is a more computationally efficient method with a running time proportional to the number of sites times the number of bonds. 

Here are a few more details of the pebble game. In a network extracted from a frictionless particle packing, since each site has 2 local degrees of freedom and there are 3 global degrees of freedom, one plays the (2,3) pebble game. Initially, there are two pebbles on each site, then these pebbles are assigned/covered to bonds one by one based on specific rules. The rules stem from an alternate version of the Laman condition, namely that the bonds in the network are independent from each other if and only if for each bond, the network formed by quadrupling the bond has no induced subnetwork of $k$ sites and greater than $2k-3$ bonds. With this reformulation, one can check when a new bond is added to the existing set of independent bonds is itself independent via quadrupling the bond in question and invoking the Laman condition. To do this, the pebble game quadruples the new bond and tries to find a pebble covering for the 4 new bonds. If a pebble covering is not found, the new bond is not an independent constraint from the others. More specifically, the pebble game is as follows:\\
\begin{enumerate}
\item Start with a set of covered bonds and add a new bond.
\item Look at the sites emanating from the new bond. If any of those sites has a free pebble, use it to cover the bond. Give a direction to this bond such that it points away from the site that has given up the pebble. Continue with another copy of the new bond. If the pebbles of the neighboring sites already cover existing bonds, then search for free pebbles in the directed network of existing edges. Once a free pebble is found, swap pebbles and reverse the arrows on the bonds  appropriately, so that the new bond is covered. Repeat this three more times. If a free pebble is found for each of the 4 copies (the quadrupled bond), then remove three of the copies and retain one bond (with its pebble and its direction) since it is added to the existing set of independent bonds.  If no free pebble is found for any of the four copies, then the new bond is not independent of the current set and it is not added to the independent set of bonds.
\item Once all the bonds in the network have been tested,  if $2N-3$ independent bonds are found, then the network is minimally rigid. If there are less than $2N-3$ independent bonds and no free pebbles, the network is overconstrained, or simply rigid, and if there are less than $2N-3$ independent bonds and free pebbles, the network is underconstrained, or floppy.
\end{enumerate}

To identify rigid clusters in the network, one introduces a new cluster label for an unlabeled bond and gathers three pebbles at its two incident sites. Then, three free pebbles are temporarily pinned down and the two incident sites marked as rigid. For each new nearest neighbor site (to the two incident sites), a pebble search is performed. If a free pebble is found, the nearest neighbor site is not mutually rigid with respect to the initial bond nor is any other site that was encountered during a pebble re-arrangement, all these sites are floppy with respect to the initial bond. However, if a free pebble is not found, the site is mutually rigid with respect to the initial bond as well as all other sites that make up the failed pebble search and so these sites are marked as rigid. Then the next-nearest neighboring sites are visited until all nearest neighbors to the set of rigid sites have been marked floppy.
All bonds between pairs of sites marked as rigid are given the same cluster label. Finally, floppy and rigid marks are removed from all sites (since a site is not unique to a rigid cluster) and the process continues until there are no unlabeled bonds. In mapping out the rigid clusters, there will be two types of bonds: isostatic bonds and redundant bonds. Isostatic bonds are critical for maintaining the rigidity of the cluster, while redundant bonds can be removed without changed the overall rigidity. Only the redudant bonds can carry stress. 

For the frictional case, we must incorporate the additional rotational degree of freedom for each particle into the pebble game. In addition, to account for the additional constraints due to tangential forces in the frictional case, we introduce a second bond for each frictional contact into the network. The pebble game then explores the network to see if that additional rotational degree of 
freedom can be independently constrained. This second bond in the network is only added to frictional contacts below the Coulomb threshold, i.e. where the normal and tangential forces are independent of each other. For contacts at the Coulomb threshold, the tangential and normal forces are no longer independent so that only one bond in the network is needed. We, therefore, arrive at a (3,3) pebble game where contacts below the Coulomb criterion are denoted as double bonds in the network and contacts at the Coulomb threshold are denoted as single bonds in the network. Two very simple networks were discussed earlier.

\section{General hierarchical lattices \label{sec:general_hierarchical}}

\begin{figure}[h]
\centering
    \includegraphics[width=0.85\columnwidth]{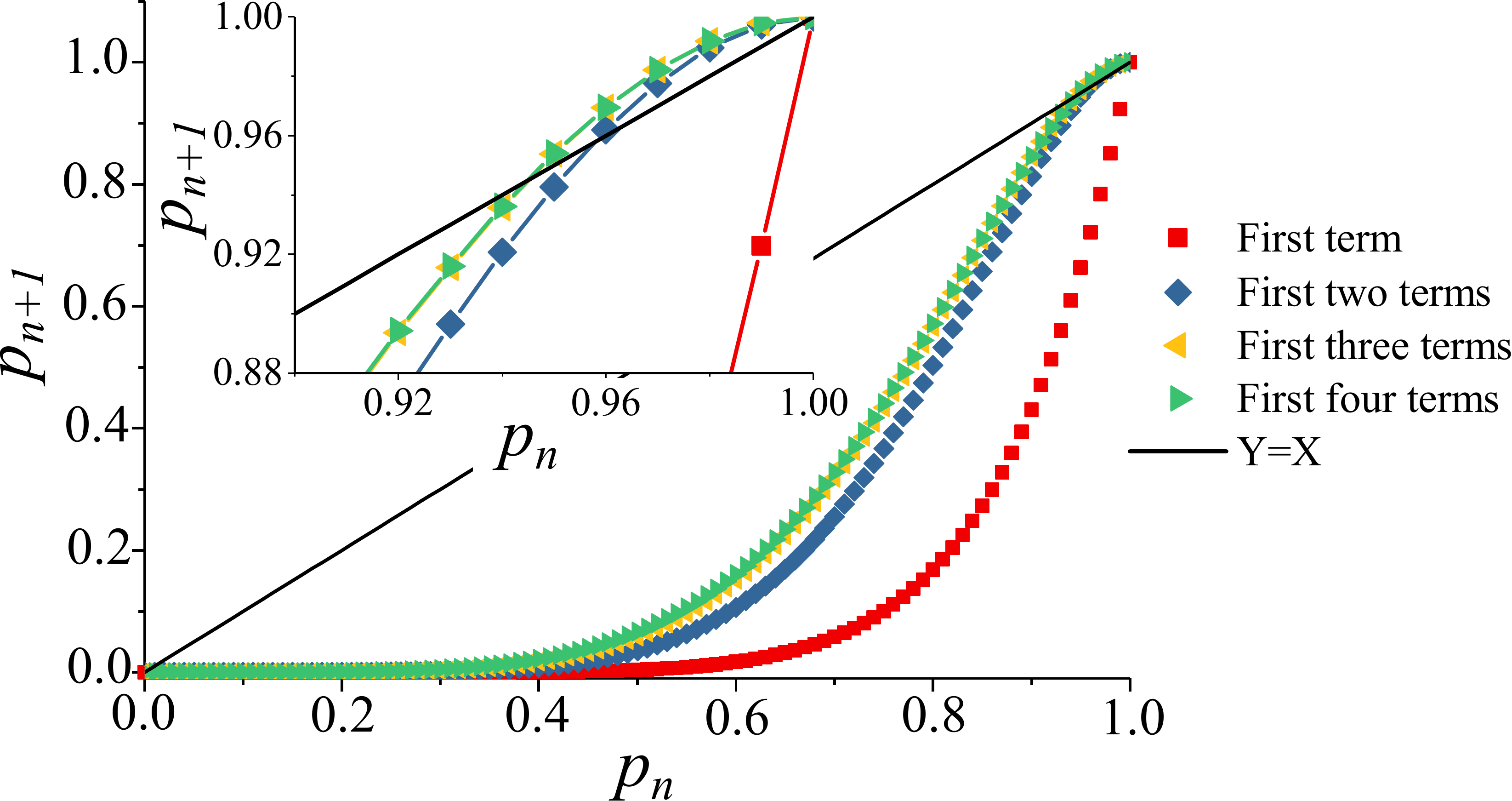}
    \caption{First four $p_n$ as function of $p$ for the $(3,3)$ game on general hierarchical lattices; $p_n$ needs at least the first two terms to generate a crossing point.}
    \label{fig:diff_orders}
\end{figure}

Let us define a basic network motif with $N_B$ bonds in a hierarchical lattice as the general first generation network and denote it by $G_0$. Then perform the subnetwork counting, like in Fig. \ref{fig:hierarchical_23}b, and assume that we find $a_n$ rigid subnetworks which have $n$ bonds less than the full network motif $G_0$. Then, generically, the recurrence relation between two generations is 
\begin{equation}
p_{n+1}=\sum_{n=0}^{N_B} a_np_n^{N_B-n}(1-p_n)^n
\label{eq:general_hierarchical}
\end{equation}
Usually $a_0=1$, and in the specific case we discussed before, $a_1=8$, $a_2=6$, $a_3=2$, and the others are zeros. The critical point is determined by the crossing point of plots of Eq. \ref{eq:general_hierarchical} and $p_{n+1}=p_n$. In Fig. \ref{fig:diff_orders}, we can see that we need at least the first two terms of Eq.~\ref{eq:general_hierarchical} to obtain a crossing point, and that they contribute much more than other terms in determining critical point $p_c$.
By same method, we can obtain $\lambda_1$ by taking derivative of Eq.~\ref{eq:general_hierarchical}, or 
\begin{equation}
\lambda_1=\sum_{n=0}^{N_B}a_n(N_B-n)p_c^{N_B-n-1}(1-p_c)^n-a_nnp_c^{N_B-n}(1-p_c)^{n-1} ,
\label{eq:general_lambda1}
\end{equation}
and use $d_f\nu=\log(N_B)/\log(\lambda_1)$ to find $d_f\nu$. The following table lists how $p_c$, $\lambda_1$ and $d_f\nu$ change when we add higher order terms to first two terms in Eq. \ref{eq:general_hierarchical}:
\begin{center}
 \begin{tabular}{||c c c c||} 
 \hline
 number of terms & $p_c$ & $\lambda_1$ & $d_f\nu$ \\ [0.5ex] 
 \hline\hline
 2 & 0.9577 & 1.8290 & 3.444 \\ 
 \hline
 3 & 0.9449 & 1.8069 & 3.5149 \\
 \hline
 4 & 0.9446 & 1.8016 & 3.5323 \\
 \hline
\end{tabular}
\end{center}

Now let's investigate the first two terms in more detail: We have $p_{n+1}=p^{N_B}+a_1p^{N_B-1}(1-p)$. To obtain a critical point, we require that
\begin{equation}
p_c^{N_B}+a_1 p_c^{N_B-1}(1-p_c)=p_c.
\label{eq:first_two}
\end{equation}
To make this equation solvable in the range $[0,1]$, we can rewrite it as: 
\begin{equation}
a_1p_c^{N_B-2}=\frac{1-p_c^{N_B-1}}{1-p_c}=1+p_c+p_c^2+...+p_c^{N_B-1}.
\label{eq:rewrited}
\end{equation}
We know $p_c$ that is a number between $0$ and $1$, so a solution requires that $a_1>N_B-1$. Since $a_1$ is the number of rigid subnetworks when just one bond is taken away from $G_0$, we have $a_1\leq N_B$. Ultimately, we obtain the equality
\begin{equation}
a_1=N_B,
\label{eq:requirement}
\end{equation}
which gives a rough criterion whether a general hierarchical lattice has a critical point, based on simple subnetwork counting.

\end{document}